\documentclass[prx,aps,nofootinbib,amssymb,twocolumn,superscriptaddress,10pt]{revtex4-2}
\usepackage[utf8]{inputenc}
\usepackage[dvipsnames]{xcolor} 
\usepackage{graphicx}
\usepackage{amsmath}
\usepackage{braket}
\usepackage{dsfont}
\usepackage{float}
\usepackage{ulem}
\usepackage{bm}

\definecolor{dodgerblue}{HTML}{1E90FF}
\definecolor{pinkerton}{HTML}{EC368D}
\usepackage[colorlinks=true,citecolor=pinkerton,linkcolor=dodgerblue,urlcolor=dodgerblue,pdftitle={moiremetals}]{hyperref}

\newcommand{\braOket}[3]{\bra{#1}#2\ket{#3}}
\newcommand{\BZ}{{\rm BZ}}
\newcommand{\Abz}{A_\BZ}
\newcommand{\Acell}{A_{\rm c}}
\newcommand{\Ham}{\mathcal{H}}

\begin{document}

\title{Moir\'e metals and continuum models beyond a single valley}
\author{Matthew Albert}
\affiliation{Department of Physics, University of Toronto, 60 St. George Street, Toronto, ON, M5S 1A7 Canada}

\author{Andrew Klekovkin-Hardy} 
\affiliation{Center for Computational Quantum Physics, Flatiron Institute, New York, New York 10010, USA}

\author{Valentin Cr\'epel}
\affiliation{Department of Physics, University of Toronto, 60 St. George Street, Toronto, ON, M5S 1A7 Canada}

\date{\today}

\begin{abstract}
We derive continuum models of moir\'e materials valid over the full monolayer Brillouin zone, as required for metals whose large Fermi surfaces are not confined to a single valley. 
To this end, we introduce an environment-resolved two-center approximation, in which each hopping amplitude depends not only on the distance between the orbitals it connects but also on its position within the moir\'e cell. 
Taking the local environment to be set by the disregistry of the two layers alone, these moir\'e couplings and their full set of harmonics can be learnt from commensurate bilayers at uniform stackings, without fitting. 
The method shows excellent agreement with ab initio band structures on moir\'e cells containing a few hundred atoms. 
Applied to twisted NbSe$_2$, it yields a continuum model in which moir\'e effects concentrate at the van Hove singularities, a single-particle starting point for studying moir\'e-tuned superconducting, charge-density-wave instabilities and their competition. 
\end{abstract}

\maketitle

\section{\label{introduction} Introduction}

The moir\'e pattern formed by the slightly mismatched lattices of superimposed two-dimensional materials modulates their electronic structure on a length scale much larger than their lattice constant. 
These moir\'e effects can stabilize phases absent in the constituent monolayers~\cite{cao2018unconventional,balents2020superconductivity,andrei2021marvels,mak2022semiconductor} and be tuned by twist angle, strain, gating, and the choice of materials~\cite{bi2019designing,kennes2021moire,nakatsuji2025high}. 
Accurately capturing this emergent long-wavelength physics proves computationally costly, because moir\'e unit cells can contain thousands of atoms at small angles~\cite{uchida2014atomic,zhang2024polarization,mao2024transfer,zhu2026twisted}. As a result, density functional theory (DFT) calculations, while essential as benchmarks, become impractical as a working tool, especially since each twist, strain, or gating configuration of interest demands a separate calculation.

Much progress in the field has circumvented this problem through the use of continuum models~\cite{lopes2007graphene,bistritzer2011moire,de2011topologically,moon2012energy}, which exploit the separation between lattice and moir\'e scales to describe the essential moiré physics with a few low-energy states coupled by smoothly varying moir\'e-periodic fields. 
These models are built around a single valley: the monolayer dispersion and the moir\'e couplings are both expanded about one high-symmetry point, with their amplitudes fixed from commensurate bilayer properties~\cite{jung2014ab,wu2019topological} or fitted to large-scale calculations~\cite{devakul2021magic,jia2024moire}.
The track record of continuum models includes the prediction of flat bands in twisted bilayer graphene~\cite{bistritzer2011moire,cao2018unconventional} and fractional Chern insulators in twisted MoTe$_2$~\cite{crepel2023anomalous,cai2023signatures,zeng2023thermodynamic}, or explaining superconductivity in twisted WSe$_2$~\cite{fischer2024theory,munoz2025twist,guo2025superconductivity,xia2025superconductivity}.

Although metals account for roughly $30\%$ of the stable monolayers~\cite{haastrup2018computational,gjerding2021recent}, their moir\'e structures still lack a continuum description, because single-valley models cannot capture their full Fermi surface. 
In the group-V TMDs NbSe$_2$ and TaS$_2$, for instance, the Fermi surface comprises pockets around both $\Gamma$ and $K$, and the van Hove singularities that control their superconducting and charge-density-wave instabilities lie about halfway along $\Gamma$--$K$~\cite{rice1975new,ugeda2016characterization,xi2016ising,nakata2018anisotropic}, far from any high-symmetry point about which a low-order expansion could be developed. 
This is in part why their twisted bilayers have received less attention, even though they are expected to host competing moir\'e-tunable instabilities~\cite{cheung2024coexisting}.

Two approximations, sometimes implicit, underlie this limitation. 
Both can be understood from a single object: the hopping coefficient between two orbitals on the moir\'e scale, $t(\delta, r_c)$, which depends on the bond length through the relative displacement $\delta$, but also on where the bond lies in the moir\'e cell -- its local environment -- through the bond center $r_c$. 
First, the two-center approximation~\cite{slater1954simplified,bistritzer2011moire} discards the dependence on $r_c$: moir\'e effects then arise from lattice interference alone, which produces interlayer tunneling but no intralayer potential. 
This is adequate for twisted bilayer graphene, where the intralayer modulation is weak~\cite{watson2023bistritzer,quinn2025higher,kang2023pseudomagnetic,vafek2023continuum}, but not for TMDs such as the ones we aim to study, where the intralayer potential is necessary and has to be supplied separately by fitting commensurate bilayers at a few high-symmetry stackings~\cite{jung2014ab,wu2019topological,zhang2021electronic,angeli2021gamma,5zt2-scbg}. 
Second, the $k\cdot p$ expansion truncates the dependence on $\delta$, keeping the monolayer dispersion to lowest order about a valley and the couplings to their value at that momentum. This is what ties current models to a single valley. 

Both truncations have since been relaxed in pursuit of better quantitative agreement with large-scale DFT band structures. 
These comparisons revealed the importance of the momentum dependence~\cite{crepel2024chiral,zhang2025twist,espitia2025quasi} and of higher harmonics~\cite{jia2024moire,mao2024transfer} of moiré couplings.  Treating them as additional fitting parameters and relying on additional physical insights yielded single-valley models achieving some transferability in twist angle~\cite{zhang2025twist}, but the fits themselves are not unique, and inconsistent parametrizations across angles have been noted~\cite{jia2024moire,zhang2025twist}. 
A fitting-free alternative was recently developed~\cite{zhang2024universal}, but it still requires a large-scale DFT calculation per angle, and the resulting model is also confined to a single valley.

What moir\'e metals require is a construction that combines the advantages of these different methods -- twist-angle transferability and low computational cost -- while maintaining accuracy across the full Brillouin zone. 
We obtain one by learning the entire function $t(\delta, r_c)$ rather than truncating or fitting it. 
More precisely, we allow all hopping amplitudes to depend on the local environment at their bond center, which we call the environment-resolved two-center approximation (ERTCA), and describe the monolayer in a basis of Wannier orbitals rather than by a $k\cdot p$ expansion, so that the dependence on $\delta$ is retained across the whole Brillouin zone. 
Taking the local environment to be set by the disregistry of the two layers at $r_c$, the moir\'e couplings become harmonics of a single twist-independent function that is sampled directly by commensurate bilayers at uniform stackings. 
A family of such small-cell first-principles calculations therefore fixes every symmetry-allowed harmonic, intralayer and interlayer alike, with their full momentum dependence, without any fitting, and once for all twist angles. 
The $k\cdot p$ models of the literature are recovered as expansions of these functions about a valley.

We benchmark the method on twisted WSe$_2$ at $5.09^\circ$~\cite{devakul2021magic}, where it reproduces large-scale DFT bands over a 250~meV window spanning several moiré bands from different valleys. 
We then apply it to twisted NbSe$_2$, for which no continuum description exists, and find that the moir\'e potential acts primarily at the van Hove singularities rather than at the band edges, which agrees with our large-scale DFT calculation at $9.43^\circ$. 
In both materials, the momentum dependence of the couplings proves decisive: it sets the bandwidth, the elementary band representation, and the valley Chern numbers of the moir\'e bands, all of which change when the couplings are frozen at the valley. 
We only treat rigid twists of identical layers, but the construction only requires Wannierizable bands and will be extended to heterobilayers, and to include strain and in-plane relaxation.

The paper is organized as follows. 
Sec.~\ref{sec:continuum} introduces the ERTCA and derives the continuum model. 
Sec.~\ref{sec:wannier} fixes its couplings from commensurate-bilayer Wannier Hamiltonians.
Sec.~\ref{sec:numerics} describes the numerical implementation, the harmonic truncation, and the benchmark on twisted WSe$_2$. Sec.~\ref{sec:NbSe2} applies the method to twisted NbSe$_2$. Sec.~\ref{sec:outlook} concludes.

\section{Continuum model derivation}
\label{sec:continuum}

\subsection{Conventions and notations}

Consider a generic monolayer material. Let $a$ denote its lattice constant, $\Lambda$ its Bravais lattice, $\Acell$ the unit-cell area, $\Lambda^{*}$ the reciprocal lattice, $\BZ$ the Brillouin zone, and $\Abz=(2\pi)^2/\Acell$ the Brillouin-zone area.
These quantities define the reference (untwisted) frame. To distinguish quantities expressed in this frame, we make them carry overbars. For example, a Wannier orbital $\alpha$ located at position $\bar{\tau}_\alpha$ within the unit cell $\bar{R}\in\Lambda$ is centered at $\bar r=\bar R+\bar\tau_\alpha$. Similarly, reciprocal lattice vectors are denoted by $\bar G\in\Lambda^*$ and crystal momenta by $\bar k\in\BZ$. 

The twisted bilayer is obtained by rigidly rotating two copies of the monolayer lattice about a common atomic site by
\begin{equation}
\theta_{t/b}=\pm \theta/2 .
\end{equation}
Throughout, we use the equivalent labels $\ell=+1\equiv t$ and $\ell=-1\equiv b$ for the top and bottom layers, respectively, switching between them as convenient. 
The lattices in the rotated frames are obtained through the linear transformations 
\begin{equation}
\Lambda_\ell = L_\ell \Lambda , \quad \Lambda_\ell^* = L_\ell^{-T} \Lambda^* , \quad \BZ_\ell = L_\ell^{-T} \BZ ,
\end{equation}
where, in the case of a pure twist considered here, $L_\ell = R(\theta_\ell)$ denotes a rotation by angle $\theta_\ell$ about the axis perpendicular to the layers. In this special case, the reciprocal-space transformation reduces to $L_\ell^{-T}=L_\ell$. We nevertheless keep the former notation, as it makes the geometric structure of the construction explicit and naturally extends to strain-induced moiré systems. 
In the bilayer, Wannier states carry an additional layer label. We therefore introduce the composite index $\beta=(\alpha,\ell)$ and denote the Wannier orbital $\alpha$ in layer $\ell$ by
\begin{equation} \label{eq:defpositionket}
\ket{r,\beta}=\ket{r,(\alpha,\ell)} , \quad r = L_\ell \bar{R} + \tau_\beta , \quad \tau_\beta = L_\ell \bar{\tau}_\alpha , 
\end{equation}
where $r$ describes the physical position of the orbital in the twisted bilayer, and $\tau_\beta $ the orbital position in the rotated unit cell. 
The corresponding quasiperiodic Bloch states are defined as 
\begin{equation} \label{eq:momentumkets}
\ket{k, \beta} = \sum_{R \in \Lambda_\ell} e^{i k \cdot r} \ket{r,\beta} . 
\end{equation}

For small twist angles, the interference between the two lattices generates a long-wavelength moiré pattern. A convenient quantity for describing this pattern is the disregistry field $d(r)$, which measures the local relative displacement between the two layers.
Since the disregistry is defined over continuous space while the atomic positions it compares are discrete, we define it via its action on plane waves on the moiré scale. 
Consider the plane waves with momenta $G_\ell=L_\ell^{-T}\bar G$ originating from the same reciprocal lattice vector $\bar G$. The disregistry field is defined so that it reproduces the relative phase accumulated between these two waves: $(G_b - G_t)\cdot r = \bar{G} \cdot d(r)$. Since this relation must hold for every $\bar G$, the disregistry field is uniquely determined as
\begin{equation}
d(r) = \mathcal{L} r, \quad \mathcal{L} = L_b^{-1} - L_t^{-1}, 
\end{equation}
which for the rigid rotations considered here reduces to $\mathcal{L} = L_t - L_b$, that is $d(r) = 2 \sin(\theta/2) \, \hat{z} \times r$.
The moiré lattice is identified with the points where all these layer-transformed plane waves still constructively interfere, \textit{i.e.} vectors satisfying $\mathcal{L} r \in \Lambda$. 
This defines the moiré lattice and its reciprocal counterparts as
\begin{equation}
\Lambda_m = \mathcal{L}^{-1} \Lambda , \quad \Lambda_m^* = \mathcal{L}^T \Lambda^* , \quad \BZ_m =  \mathcal{L}^T \BZ .
\end{equation}
Its characteristic length scale is set by the moir\'e lattice constant $a_m = a / |2\sin(\theta/2)| \gg a$ in the limit of small angles. The moir\'e unit-cell area reads $A_{\rm m} = A_{\rm c} (a_m/a)^2$.

The disregistry field gives the correspondence between atomic and moiré periodicities. For any $a_m\in\Lambda_m$ and $g\in\Lambda_m^*$, we observe that
\begin{equation}
d(r+a_m)=d(r)+\mathcal L a_m, \quad g\cdot r=\bar{G}_g\cdot d(r),
\label{eq:moiretomonolayerperiodicity}
\end{equation}
with $\mathcal L a_m\in\Lambda$ and $\bar{G}_g = (\mathcal L^{-T}g) \in\Lambda^*$. As a result, any monolayer-periodic function can be re-expanded in terms of moiré reciprocal vectors, a key ingredient of the continuum construction developed below. Throughout, moiré reciprocal vectors will be denoted by lowercase letters $g$.

\subsection{Environment resolved two-center approximation} \label{ssec:ERTCA}

Having established the necessary notation, our goal is to construct the moiré bilayer Hamiltonian $\Ham_\text{moiré}$, namely to evaluate its matrix elements $\braOket{r,\beta}{\Ham_\text{moiré}}{r',\beta'}$ in a basis of Wannier states spanning both layers. 
Unlike in an isolated crystal, these matrix elements are not determined solely by the relative displacement between the two Wannier orbitals. They also depend on the local environment in which the hopping process occurs, \textit{i.e.} they generally vary across the moiré unit cell. 
To account for this, we generalize the two-center approximation~\cite{lopes2007graphene,de2010localization,bistritzer2011moire,moon2012energy,watson2023bistritzer,quinn2025higher} and \textit{assume} that the matrix elements can be written as
\begin{equation} 
\braOket{r, \beta}{\Ham_\text{moiré}}{r', \beta'} = t_{\beta \beta'} ( \delta  , r_c ), \quad \left\{
\begin{array}{l} \delta=r'-r \\ r_c=\frac{r+r'}{2}  \end{array} \right. .
\label{eq:two-center}
\end{equation}
The hopping amplitude therefore depends not only on the relative displacement between the two orbitals, but also on the local environment sampled at their bond center. This additional dependence on the center-of-mass coordinate $r_c$ is what allows local environmental effects to be incorporated, and defines what we call the environment-resolved two-center approximation (ERTCA). The conventional two-center approximation is recovered by neglecting this dependence $t_{\beta \beta'} ( \delta  , r_c ) \to t_{\beta \beta'} ( \delta )$.

The periodicity of the moiré Hamiltonian, $\braOket{r,\beta}{\Ham_\text{moiré}}{r',\beta'} = \braOket{r+a_m,\beta}{\Ham_\text{moiré}}{r'+a_m,\beta'}$ for $a_m\in\Lambda_m$,
implies that the two arguments of $t_{\beta\beta'}(\delta,r_c)$ have distinct characters. The relative coordinate $\delta$ is unconstrained, whereas the center-of-mass coordinate $r_c$ is periodic modulo moiré lattice vectors. Consequently, the tunneling function admits a mixed Fourier representation, consisting of a continuous Fourier transform in $\delta$ and a discrete Fourier series in $r_c$:
\begin{equation} \label{eq:moire_FT} \begin{split}
t (\delta,r_c ) & = \int\frac{{\rm d}^2q}{\Abz} \sum_{g\in\Lambda_m^*} e^{-i(q\cdot\delta+ g\cdot r_c)} \, t(q, g) ,  \\   
t(q, g) & = \int \frac{{\rm d}^2\delta}{\Acell }\int \frac{{\rm d}^2r_{\rm cm}}{A_{\rm m}}  e^{i(q\cdot\delta+g\cdot r_c)}\, t (\delta,r_c) . 
\end{split}
\end{equation}
Using Eq.~\eqref{eq:moire_FT}, the momentum-space matrix elements of the moiré Hamiltonian in the basis of Eq.~\eqref{eq:momentumkets} reads as
\begin{widetext}
\begin{equation}
T_{k,k'}^{\beta, \beta'}  =  \braOket{k, \beta}{\Ham_\text{moiré}}{k', \beta'} =  \sum_{g \in \Lambda_m^*} \sum_{\substack{G_\ell \in \Lambda_\ell^* \\ G_{\ell'} \in \Lambda_{\ell'}^*}} e^{i ( \tau_\beta G_\ell - \tau_{\beta'} G_{\ell'} )}  t_{\beta \beta'} \left(\frac{ k + k' + G_\ell + G_{\ell'}}{2}; g \right) \delta_{g, (k' +G_{\ell'}) - (k + G_\ell)} . 
\label{eq:momentum_moire_tunnelings_infull}
\end{equation}
\end{widetext}
Momentum conservation $k'-k = g + G_\ell - G_{\ell'}$ naturally organizes the terms of this sum according to the magnitude of the reciprocal-lattice transfer $G_\ell - G_{\ell'}$. When this transfer is of atomic scale, momentum conservation enforces $g = G_{\ell'} - G_\ell + \mathcal{O}(a_m^{-1})$ forcing the center-of-mass harmonic $g$ to carry an atomic-scale momentum. 
In App.~\ref{app:decayharmonics}, we argue that generic smoothness and the monolayer-moiré scale separation imply an exponential asymptotic decay of $t_{\beta\beta'}(q,g)$ for $|g|\gtrsim a_m^{-1}$. Such channels are therefore  exponentially suppressed and from now on neglected. Retaining only the dominant moiré-scale scattering processes requires $G_{\ell/\ell'} = L_{\ell/\ell'}^{-T} \bar G$ to descend from the same monolayer reciprocal lattice vector $\bar G$, a property previously identified in Ref.~\cite{watson2023bistritzer}. 
Introducing the layer-direction symbol
\begin{equation}
s_{\ell\ell'} = \delta_{\ell,t}-\delta_{\ell',t} \in\{-1,0,1\},
\end{equation}
we finally obtain
\begin{align} 
T_{k,k'}^{\beta, \beta'}  & = \sum_{(g, h) \in \Lambda_m^*} e^{i(\bar{\tau}_\alpha - \bar{\tau}_{\alpha'}) \bar{G}_{h}} t_{\beta \beta'} (q^\star; g) \delta_{k'-k  , g - s_{\ell \ell'} h} , \notag \\
q^\star & = \frac{1}{2} \left\{ k + k' + \left[ L_\ell^{-T} + L_{\ell'}^{-T} \right] \bar{G}_{h} \right\} .  \label{eq:continuumERTCA}
\end{align}
Eq.~\eqref{eq:continuumERTCA} exposes two distinct sources of moiré momentum transfer. The Umklapp terms $h = [L_t^{-T} - L_b^{-T}] \bar{G}$ are purely geometric: they originate from the interference between the two transformed reciprocal lattices. These couplings survive even when the tunneling amplitudes are independent of their local environment and reproduce conventional two-center continuum models when Eq.~\eqref{eq:moire_FT} is truncated to $g=0$. 
By contrast, the center-of-mass harmonics $g\neq0$ describe a genuine modulation of the tunneling amplitudes across the moiré unit cell. They constitute the defining feature of the environment-resolved two-center approximation and generate both interlayer and intralayer moiré couplings on equal footing. In particular, the intralayer moiré terms are absent within a pure two-center approximation and arise entirely from the environmental dependence of the hopping amplitudes.

The coefficients $t_{\beta\beta'}(q;g)$ therefore provide a complete parametrization of the moiré-periodic continuum model. The next section is devoted to extracting and approximating these coefficients from small-scale first-principles calculations.

\section{Fixing the coefficients} \label{sec:wannier}
To construct the continuum model within the environment-resolved two-center approximation of Sec.~\ref{ssec:ERTCA}, we must now learn the tunneling function $t(\delta, r_c)$. 
In this section, we do so by relating them to a family of commensurate untwisted bilayer calculations.
More precisely, we consider bilayers formed from two unrotated copies of the monolayer, with the top layer rigidly translated by a displacement $d$ relative to the bottom one, the uniform disregistry of the bilayer. Since both layers share the same periodicity $\Lambda$, each $d$-shifted bilayer can be studied independently using small-scale \textit{ab initio} methods and Wannierization, yielding a family of Hamiltonians $\Ham_{\rm com}(d)$ parametrized by the relative layer displacement. 
The corresponding Wannier orbitals are labeled $\ket{\bar r, \beta=(\alpha,\ell)}$ as in Eq.~\eqref{eq:defpositionket}, and reside at position $\bar r = \bar R + \bar\tau_\alpha + \delta_{\ell,t} d$ with $\bar R \in \Lambda$.

\subsection{Registry hypothesis} \label{ssec:hypothesis}

Within the environment-resolved two-center approximation, the local environment enters through the center-of-mass coordinate $r_c$. We now make the additional assumption that this environment is fully characterized by the local relative displacement of the two layers, namely the disregistry $d(r_c)$.

To faithfully compare the tunnelings in commensurate untwisted bilayers and in the moiré heterostructure, we introduce the mean orientation of the two layers involved in the hopping process through the rotation matrix
\begin{equation}
L_{\ell\ell'} = R \left(
\frac{\theta_\ell+\theta_{\ell'}}{2} \right) = \begin{cases} L_\ell & \text{if } \ell = \ell' \\ 
 \mathds{1} & \text{if } \ell \neq \ell'
\end{cases} .
\end{equation}
$L_{\ell\ell'}$ rotates an untwisted bond to the layer frame for intralayer hoppings, and keeps it in the untwisted frame for interlayer ones, the only choice symmetric under layer exchange. 
The registry hypothesis then takes the form
\begin{equation} \label{eq:registryhyp}
t_{\beta\beta'}( \delta, r_c) = \bar t_{\beta\beta'}\left(L_{\ell\ell'}^{-1}\delta,\, d(r_c)\right) .
\end{equation}
The function $\bar t_{\beta\beta'}$ describes tunneling in the reference untwisted frame. Its first argument accounts for the orientation of the bond in the twisted structure, while its second specifies the local stacking through the disregistry. 
By construction, $\bar t_{\beta\beta'}$ is periodic over $\Lambda$ in its second argument.
Expanding $\bar{t}_{\beta\beta'}$ over its natural harmonics and using the moiré-monolayer correspondence of Eq.~\eqref{eq:moiretomonolayerperiodicity}, the Fourier coefficients satisfy
\begin{equation}
\label{eq:registryhyp_FT}
t_{\beta\beta'}(q;g) = \bar{t}_{\beta\beta'} \left(
L_{\ell\ell'}^{-1} q;\, \bar G_g \right).
\end{equation}
The continuum coefficients are therefore completely determined by the harmonics of the twist-independent function $\bar t_{\beta\beta'}$. The remainder of this section is devoted to extracting these harmonics from the commensurate bilayer family.

\subsection{Learning from commensurate bilayer}

In the commensurate bilayers, the layers are unrotated and the registry is uniform: Eq.~\eqref{eq:registryhyp} applies and Wannier matrix elements directly sample $\bar{t}_{\beta\beta'}$ directly,
\begin{equation} \label{eq:comHam}
\braOket{\bar r,\beta}{\Ham_{\rm com}(d)}{\bar r',\beta'}
= \bar{t}_{\beta\beta'}(\delta,d) ,
\end{equation}
at the separations $\delta = \bar{r}' - \bar{r} = \bar R + \bar\tau_{\alpha'} - \bar\tau_{\alpha} - s_{\ell\ell'} d$. 
The corresponding Bloch Hamiltonian reads as
\begin{equation}
W_{\beta\beta'}(\bar{k},d) = \sum_{\bar R\in \Lambda} e^{i\bar{k}\cdot\delta}\, \bar t_{\beta\beta'}(\delta,d) ,
\label{eq:WannierBlochCompact}
\end{equation}
where we retain the Fourier convention of Sec.~\ref{ssec:ERTCA}, in which the phase follows the full orbital separation $\delta$.
With this choice, $W_{\beta\beta'}(\bar k,d)$ is periodic under $d\rightarrow d+\bar a$ for any $\bar a\in\Lambda$, and therefore admits a Fourier expansion in the disregistry. The relation to the Wannier90 convention~\cite{Pizzi_2020}, where the phase depends only on the unit-cell index, is discussed in App.~\ref{app_FTconvention}.
We define the corresponding harmonics as
\begin{equation} \label{eq:Wharmonics}
W_{\beta\beta'}(\bar{k};\bar G) = \int_{\rm cell} \frac{{\rm d}^2d}{\Acell} \, e^{i\bar G\cdot d} \, W_{\beta\beta'}(\bar{k},d),
\end{equation}
with inverse $W_{\beta\beta'}(\bar{k},d)=\sum_{\bar G\in\Lambda^*}e^{-i\bar G\cdot d}W_{\beta\beta'}(\bar{k};\bar G)$. Substituting the harmonic expansion of $\bar t_{\beta\beta'}$ into Eq.~\eqref{eq:WannierBlochCompact} and performing the lattice sum yields (details in App.~\ref{app_FTconvention})
\begin{align}
W_{\beta\beta'}(\bar{k};\bar{G}) & = \! \sum_{\bar H \in\Lambda^*} \! e^{i\bar H \cdot (\bar\tau_\alpha-\bar\tau_{\alpha'})}\, \bar t_{\beta\beta'}( \bar{k}+\bar H;\, \bar{G}+\bar{H} s_{\ell\ell'}) .
\label{eq:DFTtoContinuum}
\end{align}

\subsection{Assembling the moiré Hamiltonian}

Eq.~\eqref{eq:DFTtoContinuum} provides the desired link between commensurate calculations and the moiré continuum model. 
Its left-hand side is obtained directly from the family of $d$-dependent Wannierizations, while its right-hand side contains the coefficients $\bar{t}_{\beta\beta'}(q;\bar G)$ that specify the continuum model Eq.~\eqref{eq:continuumERTCA} through the identification Eq.~\eqref{eq:registryhyp_FT}. 
The similarity between Eqs.~\eqref{eq:DFTtoContinuum} and~\eqref{eq:continuumERTCA} is immediate: the shell sum over $\bar H$ plays the same role as the harmonic sum over $h$, with $\bar H=\bar G_h$, and the accompanying phase factors coincide exactly.

The only nontrivial difference lies in the momenta $\bar{k}$ and $q^\star$ at which $\bar{t}_{\beta \beta'}$ and $t_{\beta\beta'}$ are respectively evaluated. They are related by the identity
\begin{align} \label{eq:qstar_id}
L_{\ell\ell'}^{-1} q^\star & = k^\star + \cos \left( \frac{s_{\ell\ell'}\theta}{2} \right) \bar{G}_h , \quad k^\star = L_{\ell\ell'}^{-1} \frac{k+k'}{2} ,  \notag \\
& = k^\star + \bar{G}_h + \mathcal{O}(\theta^2) .
\end{align}
Thus, up to corrections of order $\theta^2$, the momentum entering $\bar t_{\beta\beta'}$ takes the form $k^\star+\bar G_h$, with $k^\star$ independent of $h$. The shell sum in Eq.~\eqref{eq:continuumERTCA} therefore matches that of Eq.~\eqref{eq:DFTtoContinuum}, yielding
\begin{equation} \label{eq:master}
T_{k,k'}^{\beta\beta'} = \sum_{g\in\Lambda_m^*} W_{\beta\beta'}\big(k^\star;\, \bar G_g\big)\ \delta_{k'-k,\, g} .
\end{equation}

Eq.~\eqref{eq:master} shows that momenta are coupled only through moiré reciprocal vectors, while the amplitude of each channel is determined by a single registry harmonic of a commensurate-bilayer Wannier Hamiltonian. For intralayer blocks, the momentum is evaluated at $k^\star=L_\ell^{-1}\frac{k+k'}{2}$ thereby returning it to the frame of the corresponding layer and automatically reproducing the rotated monolayer dispersion. Hermiticity and layer-exchange symmetry are manifest since $L_{\ell\ell'}=L_{\ell'\ell}$. Moreover, because $W_{\beta\beta'}(\bar k;\bar G)$ is known throughout the monolayer Brillouin zone, the continuum model retains the full momentum dependence of the commensurate Wannier Hamiltonians rather than reducing the couplings to a few values near high-symmetry points.

\section{Numerical implementation}\label{sec:numerics}
This section implements the construction of Secs.~\ref{sec:continuum} and~\ref{sec:wannier}. Although it applies to any bilayer material with Wannierizable bands, we illustrate it using transition-metal dichalcogenides MX$_2$, including both semiconducting group VI (M = Mo, W) and metallic group V (M = Nb, Ta) compounds. In all cases, we obtain the continuum model without any fitting procedures  by sampling the registry and Wannierizing first-principles band structures (Sec.~\ref{ssec:Band_wannerization}). We then truncate the harmonic expansion of Eq.~\eqref{eq:master} in $\bar G$ and examine its effects on the symmetries commonly assumed in $k\cdot p$ continuum models (Sec.~\ref{sec:Trunc_symmetries}). Finally, we benchmark against large-scale DFT calculations for
twisted WSe$_2$ at $\theta = 5.09^\circ$, and find excellent agreement over  an energy window of approximately $250$~meV, where our model captures bands originating from several valleys
(Sec.~\ref{sec:large-scale_DFT}).

\begin{figure}
    \centering
    \includegraphics[width=\columnwidth]{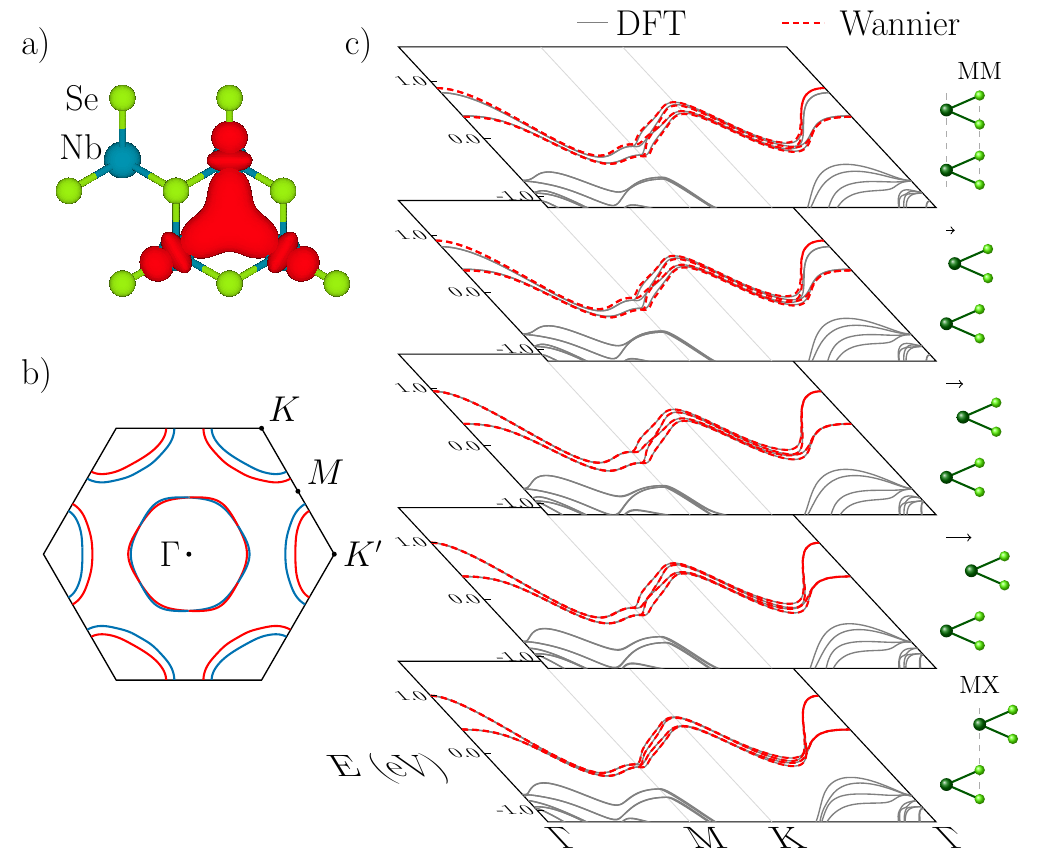} \caption{Electronic structure of NbSe$_2$. (a) Real-space lattice structure of 1H-NbSe$_2$, with the red isosurface corresponding to the $A_1$ orbital Wannier function centered at the $1c$ Wyckoff position (see App.~\ref{sec:code} for details). (b) Fermi surface of monolayer NbSe$_2$ within the first Brillouin zone, where the colors correspond to different spin bands. Ising spin-orbit coupling causes the large spin splitting near $K$ and vanishes along $\Gamma$--$M$, leading to crossings. (c) DFT and Wannier-interpolated band structure along a high-symmetry path for different stacking configurations of bilayer NbSe$_2$ (depicted schematically as inset on the right), each plotted relative to its own Fermi energy.}
    \label{fig:Fermi_Surface}
\end{figure}

\subsection{Bands and Wannierization}
\label{ssec:Band_wannerization}
The semiconducting compounds considered are classified as obstructed atomic insulators~\cite{jiang20262d}, as the C$_3$ eigenvalues of the topmost valence band at $K$ enforce an $A_1$ single-orbital representation (one per spin) at the empty $1c$ Wyckoff position located at the center of the hexagon where no atoms reside~\cite{ding2025cyclic,ingham2026real}. In the metallic compounds, the two bands near the Fermi level are isolated, half-filled, and share the same $A_1@1c$ description~\cite{yu2026quantum,cualuguaru2026observation}; we call them obstructed atomic orbitals. They are depicted for NbSe$_2$ in Fig.~\ref{fig:Fermi_Surface}(a). In
both families, Ising spin-orbit coupling splits the two spin species strongly near $K$, whereas the vertical mirror containing $\Gamma$--$M$ keeps them degenerate along this line, as visible in the monolayer spin-resolved Fermi surface of NbSe$_2$ in Fig.~\ref{fig:Fermi_Surface}(b).

We sample the registry dependence using a family of commensurate bilayers obtained from the parallel MM stacking by rigidly translating the top layer by $d$ relative to the bottom layer, followed by a full relaxation along $z$ at a fixed in-plane lattice constant (Fig.~\ref{fig:interlayer_distance}; details in App.~\ref{sec:code}). In our simulations, the translations span a uniform $12\times12$ grid of the monolayer unit cell,
\begin{equation}
  d = \frac{n}{12}\,a_1 + \frac{m}{12}\,a_2, \qquad n,m = 0,1,\dots,11,
  \label{eq:22}
\end{equation}
which contains the three high-symmetry stackings MM $(n=m=0)$, MX $(n=4, m=8)$ and XM $(n=8, m=4)$, labeled by the vertically aligned atoms of the top and bottom layers (M: metal, X:
chalcogen) as shown in Fig.~\ref{fig:Fermi_Surface}(c). We Wannierize the four bands of interest onto layer-polarized
obstructed atomic orbitals, one per spin and layer, centered at the $1c$
position of the corresponding layer. The harmonics $W_{\beta\beta'}(\bar k;\bar G)$ are meaningful only if the gauge of these independent Wannierizations varies smoothly with $d$. To ensure this, we initialize the maximal localization with projections~\cite{munoz2025twist} onto the same $A_1$ trial orbitals for every $d$. The resulting Wannier functions have spreads of order $a^2$ ($\approx 11$ \AA$^2$ for NbSe$_2$) and accurately interpolate the DFT bands (Fig.~\ref{fig:Fermi_Surface}(c)). This interpolation provides $W_{\beta\beta'}(\bar{k},d)$ of Eq.~(\ref{eq:WannierBlochCompact}) on the full grid, in the
convention of App.~\ref{app_FTconvention}. The harmonics $W_{\beta\beta'}(\bar k;\bar G)$  of Eq.~\eqref{eq:Wharmonics} are then obtained using the pseudo-inverse to maximize the accuracy.

\subsection{Truncation and recovering symmetries}
\label{sec:Trunc_symmetries}
Conventional $k\cdot p$ continuum models for the $K'$ valley retain only the three leading harmonics of the interlayer tunnelings, corresponding to the reciprocal lattice vectors $\bar{G}_{0}$, $\bar{G}_2$, and $\bar{G}_3$, where $\bar{G}_{0}=(0,0)$ and $\bar{G}_1 = (4\pi)/(\sqrt{3}a) \hat{y}$, and the remaining vectors $\bar{G}_j$ are generated by a counterclockwise rotation by angles $(j-1)\pi/3$~\cite{wu2019topological}. This truncation only preserves C$_3$ rotational symmetry when the physics is projected near one valley. It does not satisfy the symmetry requirements across the Brillouin zone. For instance, the same continuum model at C$_3 K'$ involves $\bar{G}_0$, $\bar{G}_4$, and $\bar{G}_5$. Similarly, near $\Gamma$, a faithful description requires $\bar{G}_0,...,\bar{G}_6$ from the first shell~\cite{zhang2021electronic}. More generally, the retained set of $\bar{G}_j$ must be closed under all symmetries of the commensurate family, including the point group and time reversal. Preserving C$_3$ symmetry requires truncation shell by shell; mathematically, this forces the coefficients to satisfy  
\begin{equation}
\label{eq:C3_condition}
    W_{\beta\beta'}(\bar{k},\bar{G})=W_{\beta\beta'}(\text{C}_3\bar{k},\text{C}_3\bar{G}),
\end{equation}
for all $\bar{k} \in \mathrm{BZ}$.

\begin{figure}
    \centering  \includegraphics[width=\linewidth]{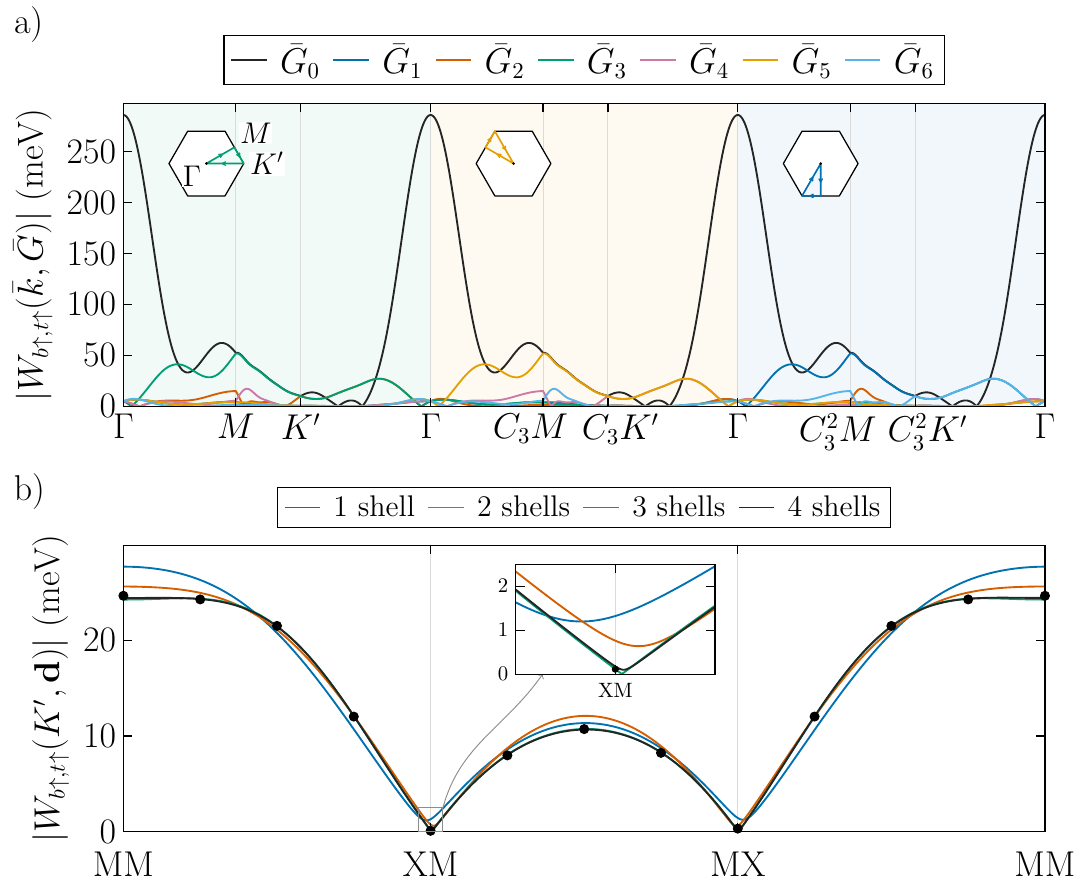}
    \caption{Magnitudes of spin-conserving interlayer tunneling harmonics in bilayer NbSe$_2$. (a) Coefficients  $W_{b\uparrow, t \uparrow}(\bar{k},\bar{G})$ along a high-symmetry path in momentum space. Note that the subleading harmonics changes in different parts of the atomic BZ. (b) Interlayer Bloch Hamiltonian matrix element $W_{b\uparrow, t \uparrow}(K',d)$ evaluated at $K'$ as a function of stacking configuration for different numbers of shells included in the expansion. Black dots denote the values obtained directly from Wannier interpolation. The inset zooms in near XM, where at least three shells are needed for the interlayer tunneling to vanish within the accuracy of the underlying DFT calculations.
    }
 \label{fig:Interlayer_tunnelings}
\end{figure}

Yet, this shell truncation, while preserving C$_3$ on the full model, does not necessarily reproduce the emergent low-energy symmetries of the $k\cdot p$ continuum models. To see this, notice that when restricted to the first shell of harmonics, the set of reciprocal lattice vectors is not closed. For example, at $K'$
\begin{equation}
\begin{aligned}
W_{\beta,\beta'}(\text{C}_3K',\text{C}_3\bar{G}_6)
  &= W_{\beta,\beta'}(K'+\bar{G}_2,\bar{G}_2) \\
  &= W_{\beta,\beta'}(K',2\, \bar{G}_2).
\end{aligned}
\end{equation}
 where $2\, \bar{G}_2$ falls outside of the first harmonic shell. Therefore, maintaining C$_3$-symmetric tunnelings requires preserving a sufficient number of shells to satisfy this condition adequately. In Fig.~\ref{fig:harmonic_decay}, we plot the dominant interlayer coefficients at the $K'$ and $\Gamma$ points as a function of harmonic magnitude. The coefficients decay exponentially with $|\bar{G}|$ and plateau beyond the fourth shell; their magnitude falls below 0.1 meV, bordering the numerical precision of our DFT calculations. Therefore, in the rest of this work, we retain harmonics up to the fourth shell. To illustrate this symmetry and truncation effects, we show the dominant interlayer components for NbSe$_2$ along a high-symmetry path in Fig.~\ref{fig:Interlayer_tunnelings}(a). Importantly, the dominant interlayer harmonics associated with $\bar{G}_0$, $\bar{G}_2$, and $\bar{G}_3$ have equal magnitudes at $K'$, consistent  with conventional moir\'e continuum models~\cite{jung2014ab,bistritzer2011moire,wu2019topological}. The harmonics preserve C$_3$ symmetry, as they satisfy Eq.~(\ref{eq:C3_condition}). Consequently, at C$_3 K'$, the harmonics associated with $\bar{G}_0$, $\bar{G}_4$, and $\bar{G}_5$ have identical magnitudes, with an analogous permutation occurring at C$_3^2 K'$. 

To demonstrate that the truncation to four shells preserves the low-energy emergent C$_3$ symmetry of continuum models and accurately reconstructs the Bloch Hamiltonian $W_{\beta\beta'}(\bar{k},d)$, Fig.~\ref{fig:Interlayer_tunnelings}(b) shows the dominant interlayer tunnelings at $K'$ along a stacking path as a function of the number of shells retained. Interlayer tunneling is maximal for MM-stackings, vanishes for MX/XM-stackings with four shells (consistent with C$_3$ symmetry), and agrees with the full Wannier interpolation (dots). As shown in the inset of Fig.~\ref{fig:Interlayer_tunnelings}(b), at least three shells are needed to recover the $C_3$-enforced zero at XM to within $0.1$~meV, the precision of the underlying DFT calculations.

\subsection{Comparison with large-scale DFT}
\label{sec:large-scale_DFT}

\begin{figure}
    \centering  \includegraphics[width=\linewidth]{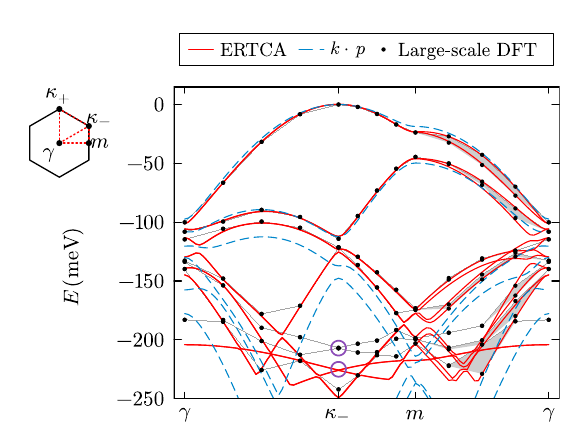}
    \caption{Benchmark of the ERTCA-derived continuum model (red lines) against existing moir\'e models for WSe$_2$ at $\theta=5.09^\circ$ twist angle. Moir\'e bands calculated through large-scale DFT data (black dots) and $k\cdot p$ continuum model around $K$ (dashed blue lines), both from  Ref.~\cite{devakul2021magic}. The grey shaded regions highlight the valley splitting in the large-scale DFT.} 
    \label{fig:Compare_large_scaleDFT}
\end{figure}

\begin{figure}
    \centering  \includegraphics[width=\linewidth]{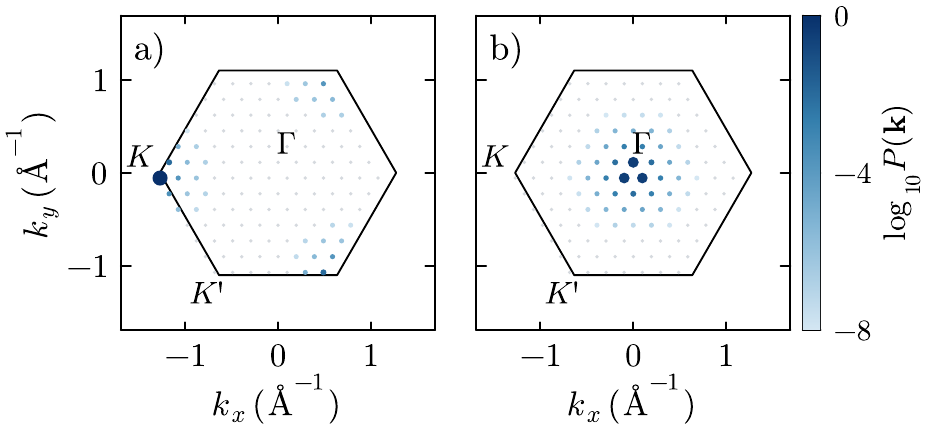}
    \caption{Valley origin of the moir\'e bands of twisted WSe$_2$ at $\theta=5.09^\circ$. Spectral weight of the moir\'e eigenstates at $q=\kappa_-$ unfolded onto the monolayer Brillouin zone, for (a) the topmost valence band and (b) the state at the band crossing near $-225$~meV (purple circle in Fig.~\ref{fig:Compare_large_scaleDFT}). For a moir\'e eigenstate $\ket{n,q}=\sum_{\beta,\,g\in\Lambda_m^*} C_{\beta,g}\ket{q+g,\beta}$, the unfolded weight is defined as $P(q+g)=\sum_\beta |C_{\beta,g}|^2$.}
    \label{fig:unfolded}
\end{figure}

To highlight our model's accuracy, which captures the momentum dependence of intra- and interlayer tunneling across the entire Brillouin zone, we compare the resulting moiré band structure with those from the conventional continuum model based on a $k\cdot p$ expansion around a high-symmetry point and the large-scale DFT spectrum of $5.09^\circ$-twisted WSe$_2$ available from Ref.~\cite{devakul2021magic}. Fig.~\ref{fig:Compare_large_scaleDFT} compares the moir\'e band structure of twisted WSe$_2$ at a twist angle of $5.09^\circ$ using our method (red lines), showing impressive agreement with the large-scale DFT calculations (black dots) over a broad 250~meV energy window.
In contrast, the $k.p$ continuum model of Ref.~\cite{devakul2021magic}, which assumes a momentum-independent intra- and interlayer tunneling fixed at the $K$ point, remains accurate within an energy window of roughly 125 meV but fails after the third moir\'e band. Crucially, our model accounts for trigonal warping effects, capturing the valley splitting (shaded in grey) observed in large-scale DFT results, which is absent in $k\cdot p$ continuum models within the effective mass approximation~\cite{wu2019topological}. 
It also resolves the band crossing at $\kappa_-$ that large-scale DFT places near $-200$~meV and our model about $25$~meV lower (purple circle in Fig.~\ref{fig:Compare_large_scaleDFT}). 
Unfolding the moir\'e eigenstates onto the monolayer Brillouin zone (Fig.~\ref{fig:unfolded}) confirms that these bands come from different valley: the topmost moir\'e valence band carries its weight near $K$ (or $K'$ for its time-reversal partner), while the state at the crossing originates from $\Gamma$. 
Such multi-valley features are natively included in our construction.

For comparison, we extract the values of the lowest harmonics in our continuum model at $K'$, which can be compared to those of $k.p$ continuum model from the literature. 
We find an intra- and interlayer tunneling amplitudes of approximately $3.9$ and $11.7$ meV, respectively. These values are significantly smaller than those from the $k\cdot p$ parametrization, which reported $9$ and $18$ meV~\cite{devakul2021magic}, and are instead closer to $2.38$ meV and $15.37$ meV extracted from large-scale DFT calculations at $3.89^\circ$~\cite{zhang2025twist}. 
One possible reason for this discrepancy is that the effective mass approximation of Ref.~\cite{devakul2021magic} breaks down in the relatively large moiré BZ at $5.09^\circ$, where trigonal warping remain non-negligible. As a result, the fitted value of $V$ and $w$ compensate for the absence of trigonal warping in the continuum model more than they describe the strength of the moiré harmonics.

\section{Moir\'e TMDs: Type-III Metals (NbSe$_2$)}
\label{sec:NbSe2}

Applying our continuum model to twisted bilayer NbSe$_2$ highlights three advantages of the method. First, its moir\'e couplings cover the entire monolayer Brillouin zone, which is necessary for metals with large Fermi surfaces, and it also parameterizes the single-valley continuum models used in the literature for all valleys at once. Second, a single calculation can quickly identify the energy and momentum regions where moir\'e effects matter most. In NbSe$_2$, these include not only band edges, the conventional focus of continuum models, but also van Hove singularities (vHS), which have received far less attention. Finally, the momentum dependence of the couplings is crucial for accurately predicting the bandwidth, elementary band representation, and topology of the moir\'e bands.

\begin{figure*}
    \centering  \includegraphics[width=\textwidth]{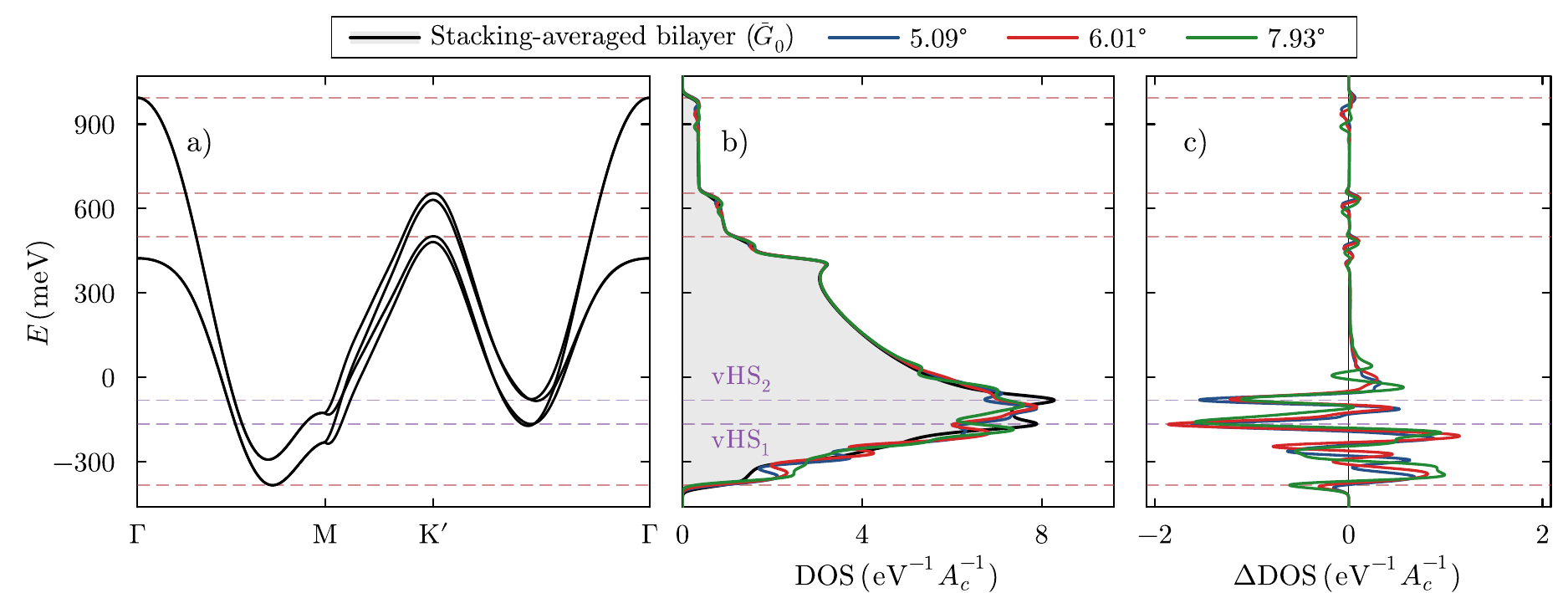}
    \caption{Density of states of twisted bilayer NbSe$_2$. (a) Stacking-averaged bilayer band structure. (b) Density of states for the stacking-averaged bilayer (black line), and the ERTCA-derived continuum model at $5.09 ^\circ$ (blue line), $6.01 ^\circ$ (red line), and $7.93 ^\circ$ (green line). (c) Change in the moir\'e system's DOS relative to the stacking-averaged bilayer. Purple horizontal dashed lines denote the two Van Hove singularities (vHS), and the red dashed lines indicate the band edges.}
    \label{fig:DOS}
\end{figure*}

\subsection{Energy resolution: where is moir\'e?} \label{ssec:where_moire}
A key advantage of our method is that it identifies, in a single pass and without fitting, all energies where moir\'e effects are relevant, providing a starting point for developing effective models near each. To locate these effects, we compute the density of states with and without the moir\'e potential.

As a reference for the case without moir\'e effects, we use the stacking-averaged bilayer, whose Hamiltonian only retains the stacking-independent coefficients $W_{\beta\beta'}(\bar k;\bar G_0)$. It can be viewed as the unperturbed bilayer before any moir\'e coupling is turned on. Fig.~\ref{fig:DOS}(a) shows this stacking-averaged band structure for bilayer NbSe$_2$, and the black solid lines in Fig.~\ref{fig:DOS}(b) display the corresponding density of states (DOS). Two prominent spin-split vHS (purple dashed lines) arise from saddle points located roughly halfway along the $\Gamma-K'$ path, also present in monolayer NbSe$_2$~\cite{nakata2018anisotropic,ingham2026real}. Band edges at $K'$ and $\Gamma$ appear as steps in the DOS (red horizontal dashed lines), as expected of nearly quadratic two-dimensional bands.

Fig.~\ref{fig:DOS}(b) also presents the DOS with the moir\'e potential included, computed as described in Secs.~\ref{sec:continuum}~-~\ref{sec:numerics} for twisted NbSe$_2$ at $\theta=5.09^\circ$, $6.01^\circ$ and $7.93^\circ$, and Fig.~\ref{fig:DOS}(c) shows its change $\Delta\mathrm{DOS}$ relative to the stacking-averaged reference.
 This change is dominated by sharp dips at the two vHS, flanked by positive lobes at nearby energies, a sign that spectral weight is redistributed away from the vHS energies. This redistribution originates from moir\'e scattering, which couples the vHS to states shifted by moir\'e reciprocal vectors; Sec.~\ref{ssec:full_BZ} provides a complementary momentum-space picture (Fig.~\ref{fig:NbSe2_spectral_weight}). Away from the vHS, weaker and more conventional moiré effects appear near the band edges (red dashed lines in Fig.~\ref{fig:DOS}). These are the
regions usually studied by $k\cdot p$ continuum models expanded around high-symmetry points~\cite{wu2019topological,bistritzer2011moire}, which our construction recovers by projection onto small momentum patches (Sec.~\ref{ssec:full_BZ}). The DOS elsewhere remains largely unaffected.

\begin{figure}
    \centering  \includegraphics[width=\linewidth]{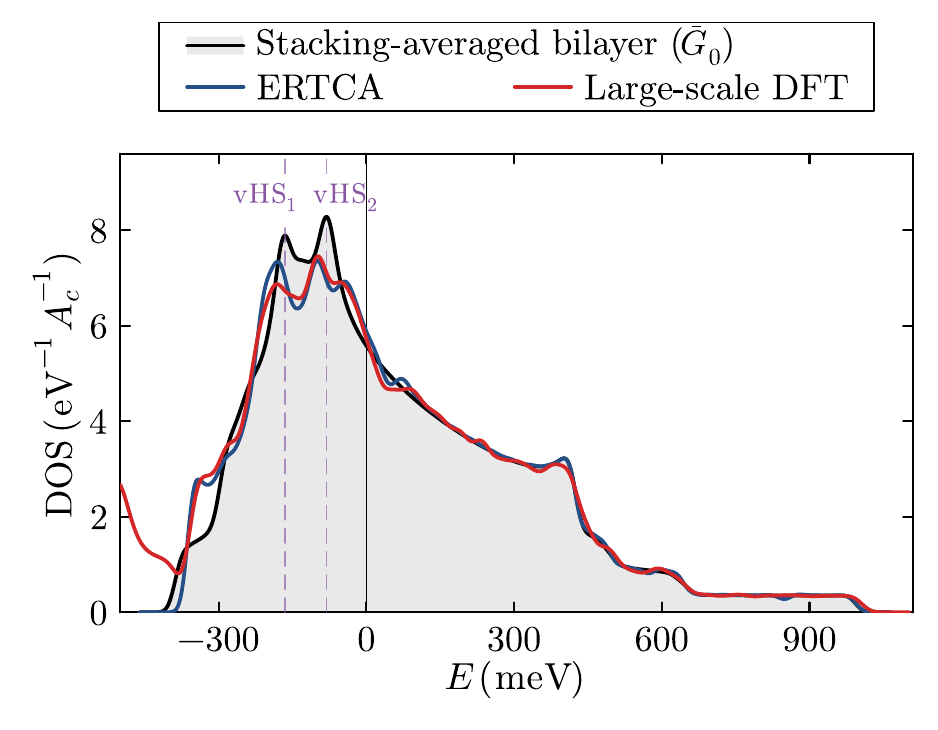}
    \caption{Comparison of the density of states of twisted bilayer NbSe$_2$ at $9.43^\circ$. The DOS for the stacking-averaged bilayer (black line), the twisted structure at $9.43 ^\circ$ obtained through large-scale DFT (red lines), and the ERTCA-derived continuum model (blue line).}
\label{fig:NbSe2_large_scale}
\end{figure}
To verify our findings, we compare it with large-scale DFT calculations performed for bilayer NbSe$_2$ at $\theta = 9.43^\circ$ (222 atoms per moiré unit cell).
Fig.~\ref{fig:NbSe2_large_scale} shows a remarkable agreement between the two
DOS over the entire bandwidth, including the predicted reduction at both vHS and weaker effects near the band edges. 
The agreement is only qualitative, as expected, since the cost of the large-scale calculation limits us to a $6\times6\times1$ self-consistent and a $12\times12\times1$ non-self-consistent $k$-grid, which only coarsely resolves the DOS and its moir\'e-scale variations. To suppress the resulting grid artifacts, the large-scale DOS is convolved with a $15$~meV Gaussian (App.~\ref{sec:code}), which precludes any quantitative comparison on smaller energy scales. 
Finally, note that we restricted our analysis to $\theta \geq 5^\circ$, safely above the threshold at which in-plane lattice relaxation sets in~\cite{nakatsuji2025high}. Such relaxation alters the local disregistry and is neglected in our rigid model; including its effects is left for future work.

\subsection{Momentum resolution: beyond a single valley} \label{ssec:full_BZ}

\begin{figure}
    \centering\includegraphics[width=\columnwidth]{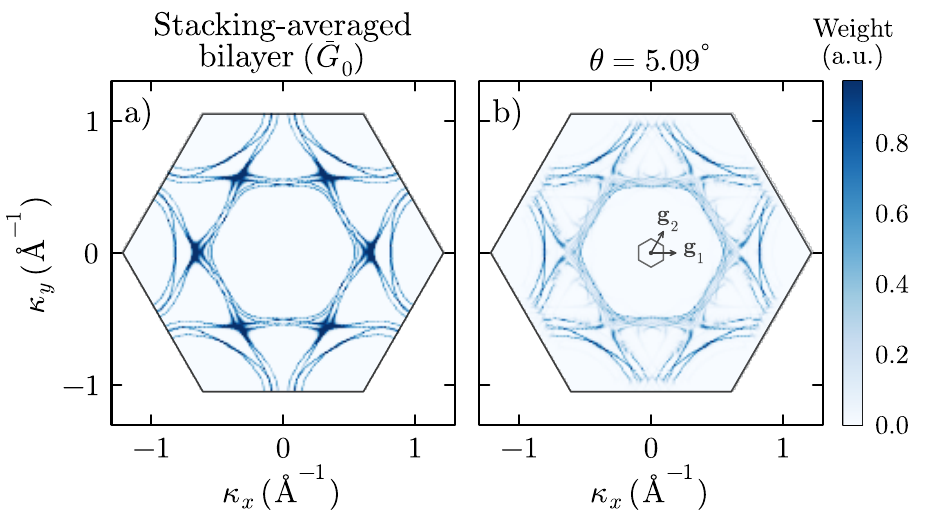}
    \caption{Momentum-resolved spectral weight of NbSe$_2$ within the first monolayer Brillouin zone at $-81$ meV relative to the Fermi energy, corresponding to the upper Van Hove singularity (vHS$_2$ in the middle panel of Fig.~\ref{fig:DOS}). Left panel corresponds to the untwisted, stacking-averaged bilayer. Right panel shows the spectral function for the twisted structure at a twist angle of $\theta=5.09^\circ$.} 
    \label{fig:NbSe2_spectral_weight}
\end{figure}

One of the main motivations of this work is to describe moir\'e metals such as twisted bilayer NbSe$_2$, whose active states near the Fermi level are spread across the monolayer Brillouin zone. Our method captures the moir\'e scattering of all these states simultaneously, as illustrated in Fig.~\ref{fig:NbSe2_spectral_weight}. It shows the momentum-resolved spectral function at the energy of vHS$_2$ (Fig.~\ref{fig:DOS}), for the stacking-averaged bilayer (a) and with the moir\'e coupling turned on at $\theta=5.09^\circ$ (b). The moir\'e reconstruction is clearly visible in the smearing of the inner hexagonal contour, the drastic loss of spectral weight at the saddle points, and the vanishing of the weight along the edge of the monolayer Brillouin zone around $M$ at this twist angle.\par
This contrasts with conventional continuum models, which are built around
a single valley: the monolayer dispersion is expanded to lowest order in
$k\cdot p$, and the moir\'e couplings are fixed to their values at the
valley momentum -- typically extracted from commensurate or large-scale
first-principles calculations~\cite{jung2014ab,bistritzer2011moire,wu2019topological,jia2024moire,
zhang2024universal}.\par
Our construction reproduces these models, for all valleys at once, by
projecting onto a small patch of the monolayer Brillouin zone, i.e., by
retaining in Eq.~\eqref{eq:master} only momenta within a few moir\'e shells
of a given high-symmetry point. This includes valleys buried deep in the
spectrum of large-scale calculations, such as the lower $\Gamma$ band of
bilayer NbSe$_2$, which lies below the maxima around $K'$ yet can be studied
independently within our framework, as shown in Fig.~\ref{fig:NbSe2_continuum}.

\subsection{Importance of momentum dependence}
\label{ssec:k_dependence}

Our method offers an unprecedented understanding of the momentum dependence of the moir\'e couplings, the $W_{\beta\beta'}(\bar k;\bar G)$ coefficients, which vary strongly across the monolayer Brillouin zone [Fig.~\ref{fig:Interlayer_tunnelings}(a)]. Conventional continuum models discard these variations, but they are essential for correctly capturing the bandwidth, the elementary band representation, and the topology of the moiré bands.

\subsubsection{Bandwidth}\label{sssec:bandswidth}

For twisted WSe$_2$, the bandwidth of the topmost moir\'e band at $\theta=5.09^\circ$ is set mostly by the monolayer dispersion rather than by the moir\'e couplings. Without moir\'e couplings, the folded bands of the decoupled, rotated layers already have a width $E(\kappa_+) - E(\gamma) = 119.9$~meV, while the moir\'e couplings reduce it to $102.1$~meV, in close agreement with the $100.1$~meV of large-scale DFT~\cite{devakul2021magic} (Fig.~\ref{fig:Compare_large_scaleDFT}). Replacing the exact monolayer dispersion by its effective-mass approximation ($m^*=0.345\,m_e$) neglects trigonal warping and inflates these widths by about $20\%$, to $140.5$~meV and $122.4$~meV, respectively. A quadratic model fitted to the large-scale DFT bands must therefore compensate for this excess kinetic width with artificially large moir\'e couplings. This explains why the parameters of Ref.~\cite{devakul2021magic} ($9$ and $18$~meV for the intra- and interlayer couplings) are significantly larger than both those extracted from large-scale DFT at $3.89^\circ$ and validated across $2-5^\circ$ ($2.38$ and $15.37$~meV~)\cite{zhang2025twist} and ours evaluated at $K'$ ($3.9$ and $11.7$~meV), the latter two agreeing on a much weaker intralayer potential.

Beyond corrections to the dispersion, the momentum dependence of the couplings themselves becomes essential in bilayer NbSe$_2$. As seen in Fig.~\ref{fig:Interlayer_tunnelings}(a), the dominant interlayer harmonic $\bar G_0$ drops from about $286$~meV at $\Gamma$ to below $50$~meV over a momentum scale comparable to the size of the moir\'e Brillouin zone at $5.09^\circ$ [Fig~\ref{fig:NbSe2_spectral_weight}(b)]. 
Fixing the interlayer tunneling to its $\Gamma$-point value therefore overestimates the coupling by a factor of $3$ to $5$ or more almost everywhere, directly affecting the predicted moir\'e bandwidth, as Fig.~\ref{fig:NbSe2_continuum} illustrates for the bands around $\Gamma$. 
Retaining four harmonic shells in all cases, the continuum model with couplings fixed to their $\Gamma$-point values (blue dashed lines) predicts a bandwidth of about $70$~meV for the highest moir\'e bands, compared with $160$~meV for our momentum-dependent model (solid red lines). 
Most of this discrepancy comes from the constant $\bar G_0$ harmonic: incorporating momentum dependence in this harmonic alone increases the bandwidth to nearly $140$~meV (black dashed lines).

\subsubsection{Elementary band representations}

\begin{figure}
    \centering  \includegraphics[width=\linewidth]{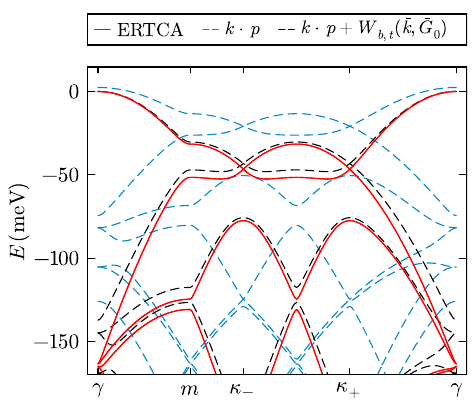}
    \caption{NbSe$_2$ moir\'e bands at $\theta=5.09^\circ$ twist angle around $\Gamma$ including four harmonic shells calculated through $k\cdotp$ continuum model (dashed blue) and $k\cdot p$ with momentum-dependent interlayer tunneling that corresponds to $\bar{G}_0$ (black dashed).} 
    \label{fig:NbSe2_continuum}
\end{figure}

In Fig.~\ref{fig:NbSe2_continuum}, the top two $k\cdot p$-derived moir\'e bands form a doublet at $\kappa_{\pm}$ and non-degenerate singlets at $\gamma$, consistent with an $s$-orbital on a hexagonal lattice~\cite{bradlyn2017topological,cano2018building}. With momentum-dependent couplings (solid red lines), the highest bands instead form an isolated set triplet, with a singlet and a doublet at both $\gamma$ and $\kappa_\pm$, consistent with an $s$ orbital on a kagome lattice. 

Previous studies of $\Gamma$-valley TMD homobilayers~\cite{angeli2021gamma,xian2021realization} found honeycomb-lattice physics in the highest bands and kagome physics only deep in the spectrum, where Hartree-Fock corrections are expected, and at small angles ($\theta\lesssim3^\circ$), where lattice relaxation may eliminate it~\cite{angeli2021gamma,xian2021realization}. In our model, retaining four shells, momentum-dependent tunneling, and no in-plane lattice relaxations on the moir\'e scale, we find a transition at roughly {$\theta = 3.4^\circ$}, where the first set of bands changes from a Kagome lattice to a hexagonal lattice. These differences alter the basis choice when Wannierizing the continuum model.

\subsubsection{Topology}
The topological nature of the first moiré valence bands in twisted WSe$_2$ has been debated. Theoretical works have predicted both trivial~\cite{nakatsuji2025high,pan2020band,brzezinska2024pressure} and topological~\cite{devakul2021magic} topmost valence moir\'e bands, with topology sensitive to the twist angle~\cite{zhang2024polarization,zhang2025twist} and displacement fields~\cite{hu2022nonlinear}. Experiments, on the other hand, have only observed topological moir\'e bands at small twist angles ($1.2^\circ-3^\circ$)~\cite{foutty2024mapping,kang2024double,gao2026probing}, and physics compatible with a topologically trivial triangular lattice at larger angles ($\theta\gtrsim3.5^\circ$)~\cite{wang2020correlated,guo2025superconductivity,guo2025superconductivity,xia2025superconductivity,xia2026bandwidth,xia2026extended,fischer2024theory,munoz2025twist}.

We find that capturing the momentum dependence of the tunnelings is key to accurately describing their band topology. For twisted WSe$_2$ at $5.09^\circ$ (Fig.~\ref{fig:Compare_large_scaleDFT}), our method predicts that the highest energy band, originating from the $K$ valley, has a valley Chern number $C_{K}=-1$, while the second highest valley-split band (shaded grey) has C$_{K'}=1$. These nonzero valley Chern numbers agree with large-scale DFT calculations performed at the same angle~\cite{devakul2021magic,zhang2025twist}. 
When removing the momentum dependence of moiré coupling by evaluating all moiré coefficients of our model at the valley momentum except for the intralayer $\bar{G}_0$ terms making up the stacking-averaged reference, we instead find that both the same bands are driven to a topologically trivial state with $C_{K}=C_{K'}=0$. 
This reproduces predictions from earlier $k\cdot p$ continuum models for twisted bilayer WSe$_2$~\cite{pan2020band,nakatsuji2025high} and shows that the momentum dependence of the moir\'e couplings is essential for correctly identifying band topology in large-angle twisted WSe$_2$. 
We also note that the continuum model in Ref.~\cite{devakul2021magic}, obtained with a different fitting procedure (Sec.~\ref{sssec:bandswidth}), finds topological moir\'e bands at the $k\cdot p$ level.

\section{Outlook} \label{sec:outlook}

In this work, we developed a continuum model for moiré materials that applies to monolayers with large Fermi surfaces, such as NbSe$_2$. Our fitting-free approach achieves full Brillouin zone accuracy comparable to large-scale DFT calculations for moir\'e unit cells containing a few hundred atoms ($\theta \gtrsim 5^\circ$), while maintaining the computational efficiency of conventional single-valley continuum models. The key to our method lies in learning the momentum dependence of the moir\'e couplings, which in real space corresponds to parametrizing the variations of the Hamiltonian across the moir\'e unit cell. We formalized this through an environment-resolved two-center approximation, a generalization of the conventional two-center approximation~\cite{slater1954simplified,bistritzer2011moire} in which each hopping amplitude depends on both the relative displacement of the two orbitals and the local environment at their bond center. This environment dependence is the minimal addition that generates the intralayer moir\'e potentials missing in the pure two-center treatment.
The momentum dependence of the moir\'e couplings proves critical to capture the moir\'e-reconstructed Fermi surface, as well as the bandwidth, elementary band representation, and topology of the moir\'e bands.

The construction can be extended beyond the present rigid-twist limit considered throughout. Both in-plane lattice relaxation and strain can be included by changing the map from moir\'e coordinates to local environment -- in this work, assumed to be fully characterized by the disregistry field $d(r)$ -- while keeping the commensurate coefficients unchanged. Relaxation replaces the rigid disregistry by a relaxed one, and heterostrain modifies the linear maps $L_\ell$, which can treat uniform strain and twist on the same footing. Heterobilayers and other stacking arrangements only require identifying the corresponding family of small-cell commensurate calculations. 

Because our method accurately describes the moir\'e bands, it provides a reliable starting point for studying interactions. By simultaneously capturing remote bands and multiple valleys, it is ideally suited to interaction-driven phases where band mixing is essential, such as the fractional Chern insulators in rhombohedral graphene~\cite{li2025multiband}. Another example is twisted NbSe$_2$: the monolayer hosts both superconductivity and a charge density wave~\cite{nakata2018anisotropic}, and twisted bilayers have been reported to host coexisting charge density waves~\cite{cheung2024coexisting}. We showed that moir\'e couplings modify the density of states in twisted NbSe$_2$ primarily near the van Hove singularities and band edges, indicating that twist can act as a knob to tune the electronic states that enter both instabilities. Our continuum model provides a natural framework to explore this tunability, by projecting interactions and pairing channels onto moir\'e states that span the entire Fermi surface.

Future extensions will capture higher-order environmental corrections using machine-learning techniques~\cite{wang2025atomic} that learn the real-space tunnelings from a limited set of DFT calculations.

\begin{acknowledgments}

We thank S. de la Barrera for discussions on moir\'e systems with large Fermi surfaces, from which this study originated.

Ce travail a été financé par le Conseil de recherches en sciences naturelles et en génie du Canada (CRSNG), subvention à la découverte RGPIN-2026-07726 et supplément de lancement à la découverte DGECR-2026-00075, ainsi que par une bourse Connaught pour nouveaux chercheurs 2025-2026 (Connaught New Researcher Award) du Fonds Connaught de l’Université de Toronto. Cette recherche a été rendue possible en partie grâce au soutien de Compute Ontario (computeontario.ca) et de l’Alliance de recherche numérique du Canada (alliancecan.ca), dans le cadre de l’allocation Ressources pour les groupes de recherche rrg-vcrepel (RRG 2026). MA est titulaire d’une Bourse d’études supérieures de l’Ontario (BESO / Ontario Graduate Scholarship, OGS). The Flatiron Institute is a division of the Simons Foundation.

\end{acknowledgments}

\section*{Data availability}

All data and code needed to reproduce the figures and findings of this work are publicly available on \href{https://github.com/matthew7879/ERTCA.git}{GitHub}.

\section*{Use of AI tools}

The authors used OpenAI ChatGPT (GPT-6 Astra) and OpenAI Codex tools for assistance with continuum model implementation and plotting. All AI-assisted code and figures were directed and independently verified by the authors, who take full responsibility for the content of this work.

\bibliography{aps}

\begin{thebibliography}{73}%
\makeatletter
\providecommand \@ifxundefined [1]{%
 \@ifx{#1\undefined}
}%
\providecommand \@ifnum [1]{%
 \ifnum #1\expandafter \@firstoftwo
 \else \expandafter \@secondoftwo
 \fi
}%
\providecommand \@ifx [1]{%
 \ifx #1\expandafter \@firstoftwo
 \else \expandafter \@secondoftwo
 \fi
}%
\providecommand \natexlab [1]{#1}%
\providecommand \enquote  [1]{``#1''}%
\providecommand \bibnamefont  [1]{#1}%
\providecommand \bibfnamefont [1]{#1}%
\providecommand \citenamefont [1]{#1}%
\providecommand \href@noop [0]{\@secondoftwo}%
\providecommand \href [0]{\begingroup \@sanitize@url \@href}%
\providecommand \@href[1]{\@@startlink{#1}\@@href}%
\providecommand \@@href[1]{\endgroup#1\@@endlink}%
\providecommand \@sanitize@url [0]{\catcode `\\12\catcode `\$12\catcode `\&12\catcode `\#12\catcode `\^12\catcode `\_12\catcode `\%12\relax}%
\providecommand \@@startlink[1]{}%
\providecommand \@@endlink[0]{}%
\providecommand \url  [0]{\begingroup\@sanitize@url \@url }%
\providecommand \@url [1]{\endgroup\@href {#1}{\urlprefix }}%
\providecommand \urlprefix  [0]{URL }%
\providecommand \Eprint [0]{\href }%
\providecommand \doibase [0]{https://doi.org/}%
\providecommand \selectlanguage [0]{\@gobble}%
\providecommand \bibinfo  [0]{\@secondoftwo}%
\providecommand \bibfield  [0]{\@secondoftwo}%
\providecommand \translation [1]{[#1]}%
\providecommand \BibitemOpen [0]{}%
\providecommand \bibitemStop [0]{}%
\providecommand \bibitemNoStop [0]{.\EOS\space}%
\providecommand \EOS [0]{\spacefactor3000\relax}%
\providecommand \BibitemShut  [1]{\csname bibitem#1\endcsname}%
\let\auto@bib@innerbib\@empty
\bibitem [{\citenamefont {Cao}\ \emph {et~al.}(2018)\citenamefont {Cao}, \citenamefont {Fatemi}, \citenamefont {Fang}, \citenamefont {Watanabe}, \citenamefont {Taniguchi}, \citenamefont {Kaxiras},\ and\ \citenamefont {Jarillo-Herrero}}]{cao2018unconventional}%
  \BibitemOpen
  \bibfield  {author} {\bibinfo {author} {\bibfnamefont {Y.}~\bibnamefont {Cao}}, \bibinfo {author} {\bibfnamefont {V.}~\bibnamefont {Fatemi}}, \bibinfo {author} {\bibfnamefont {S.}~\bibnamefont {Fang}}, \bibinfo {author} {\bibfnamefont {K.}~\bibnamefont {Watanabe}}, \bibinfo {author} {\bibfnamefont {T.}~\bibnamefont {Taniguchi}}, \bibinfo {author} {\bibfnamefont {E.}~\bibnamefont {Kaxiras}},\ and\ \bibinfo {author} {\bibfnamefont {P.}~\bibnamefont {Jarillo-Herrero}},\ }\href@noop {} {\bibfield  {journal} {\bibinfo  {journal} {Nature}\ }\textbf {\bibinfo {volume} {556}},\ \bibinfo {pages} {43} (\bibinfo {year} {2018})}\BibitemShut {NoStop}%
\bibitem [{\citenamefont {Balents}\ \emph {et~al.}(2020)\citenamefont {Balents}, \citenamefont {Dean}, \citenamefont {Efetov},\ and\ \citenamefont {Young}}]{balents2020superconductivity}%
  \BibitemOpen
  \bibfield  {author} {\bibinfo {author} {\bibfnamefont {L.}~\bibnamefont {Balents}}, \bibinfo {author} {\bibfnamefont {C.~R.}\ \bibnamefont {Dean}}, \bibinfo {author} {\bibfnamefont {D.~K.}\ \bibnamefont {Efetov}},\ and\ \bibinfo {author} {\bibfnamefont {A.~F.}\ \bibnamefont {Young}},\ }\href@noop {} {\bibfield  {journal} {\bibinfo  {journal} {Nature Physics}\ }\textbf {\bibinfo {volume} {16}},\ \bibinfo {pages} {725} (\bibinfo {year} {2020})}\BibitemShut {NoStop}%
\bibitem [{\citenamefont {Andrei}\ \emph {et~al.}(2021)\citenamefont {Andrei}, \citenamefont {Efetov}, \citenamefont {Jarillo-Herrero}, \citenamefont {MacDonald}, \citenamefont {Mak}, \citenamefont {Senthil}, \citenamefont {Tutuc}, \citenamefont {Yazdani},\ and\ \citenamefont {Young}}]{andrei2021marvels}%
  \BibitemOpen
  \bibfield  {author} {\bibinfo {author} {\bibfnamefont {E.~Y.}\ \bibnamefont {Andrei}}, \bibinfo {author} {\bibfnamefont {D.~K.}\ \bibnamefont {Efetov}}, \bibinfo {author} {\bibfnamefont {P.}~\bibnamefont {Jarillo-Herrero}}, \bibinfo {author} {\bibfnamefont {A.~H.}\ \bibnamefont {MacDonald}}, \bibinfo {author} {\bibfnamefont {K.~F.}\ \bibnamefont {Mak}}, \bibinfo {author} {\bibfnamefont {T.}~\bibnamefont {Senthil}}, \bibinfo {author} {\bibfnamefont {E.}~\bibnamefont {Tutuc}}, \bibinfo {author} {\bibfnamefont {A.}~\bibnamefont {Yazdani}},\ and\ \bibinfo {author} {\bibfnamefont {A.~F.}\ \bibnamefont {Young}},\ }\href@noop {} {\bibfield  {journal} {\bibinfo  {journal} {Nature Reviews Materials}\ }\textbf {\bibinfo {volume} {6}},\ \bibinfo {pages} {201} (\bibinfo {year} {2021})}\BibitemShut {NoStop}%
\bibitem [{\citenamefont {Mak}\ and\ \citenamefont {Shan}(2022)}]{mak2022semiconductor}%
  \BibitemOpen
  \bibfield  {author} {\bibinfo {author} {\bibfnamefont {K.~F.}\ \bibnamefont {Mak}}\ and\ \bibinfo {author} {\bibfnamefont {J.}~\bibnamefont {Shan}},\ }\href@noop {} {\bibfield  {journal} {\bibinfo  {journal} {Nature Nanotechnology}\ }\textbf {\bibinfo {volume} {17}},\ \bibinfo {pages} {686} (\bibinfo {year} {2022})}\BibitemShut {NoStop}%
\bibitem [{\citenamefont {Bi}\ \emph {et~al.}(2019)\citenamefont {Bi}, \citenamefont {Yuan},\ and\ \citenamefont {Fu}}]{bi2019designing}%
  \BibitemOpen
  \bibfield  {author} {\bibinfo {author} {\bibfnamefont {Z.}~\bibnamefont {Bi}}, \bibinfo {author} {\bibfnamefont {N.~F.}\ \bibnamefont {Yuan}},\ and\ \bibinfo {author} {\bibfnamefont {L.}~\bibnamefont {Fu}},\ }\href@noop {} {\bibfield  {journal} {\bibinfo  {journal} {Physical Review B}\ }\textbf {\bibinfo {volume} {100}},\ \bibinfo {pages} {035448} (\bibinfo {year} {2019})}\BibitemShut {NoStop}%
\bibitem [{\citenamefont {Kennes}\ \emph {et~al.}(2021)\citenamefont {Kennes}, \citenamefont {Claassen}, \citenamefont {Xian}, \citenamefont {Georges}, \citenamefont {Millis}, \citenamefont {Hone}, \citenamefont {Dean}, \citenamefont {Basov}, \citenamefont {Pasupathy},\ and\ \citenamefont {Rubio}}]{kennes2021moire}%
  \BibitemOpen
  \bibfield  {author} {\bibinfo {author} {\bibfnamefont {D.~M.}\ \bibnamefont {Kennes}}, \bibinfo {author} {\bibfnamefont {M.}~\bibnamefont {Claassen}}, \bibinfo {author} {\bibfnamefont {L.}~\bibnamefont {Xian}}, \bibinfo {author} {\bibfnamefont {A.}~\bibnamefont {Georges}}, \bibinfo {author} {\bibfnamefont {A.~J.}\ \bibnamefont {Millis}}, \bibinfo {author} {\bibfnamefont {J.}~\bibnamefont {Hone}}, \bibinfo {author} {\bibfnamefont {C.~R.}\ \bibnamefont {Dean}}, \bibinfo {author} {\bibfnamefont {D.}~\bibnamefont {Basov}}, \bibinfo {author} {\bibfnamefont {A.~N.}\ \bibnamefont {Pasupathy}},\ and\ \bibinfo {author} {\bibfnamefont {A.}~\bibnamefont {Rubio}},\ }\href@noop {} {\bibfield  {journal} {\bibinfo  {journal} {Nature Physics}\ }\textbf {\bibinfo {volume} {17}},\ \bibinfo {pages} {155} (\bibinfo {year} {2021})}\BibitemShut {NoStop}%
\bibitem [{\citenamefont {Nakatsuji}\ \emph {et~al.}(2025)\citenamefont {Nakatsuji}, \citenamefont {Cano},\ and\ \citenamefont {Cr{\'e}pel}}]{nakatsuji2025high}%
  \BibitemOpen
  \bibfield  {author} {\bibinfo {author} {\bibfnamefont {N.}~\bibnamefont {Nakatsuji}}, \bibinfo {author} {\bibfnamefont {J.}~\bibnamefont {Cano}},\ and\ \bibinfo {author} {\bibfnamefont {V.}~\bibnamefont {Cr{\'e}pel}},\ }\href@noop {} {\bibfield  {journal} {\bibinfo  {journal} {arXiv preprint arXiv:2512.15851}\ } (\bibinfo {year} {2025})}\BibitemShut {NoStop}%
\bibitem [{\citenamefont {Uchida}\ \emph {et~al.}(2014)\citenamefont {Uchida}, \citenamefont {Furuya}, \citenamefont {Iwata},\ and\ \citenamefont {Oshiyama}}]{uchida2014atomic}%
  \BibitemOpen
  \bibfield  {author} {\bibinfo {author} {\bibfnamefont {K.}~\bibnamefont {Uchida}}, \bibinfo {author} {\bibfnamefont {S.}~\bibnamefont {Furuya}}, \bibinfo {author} {\bibfnamefont {J.-I.}\ \bibnamefont {Iwata}},\ and\ \bibinfo {author} {\bibfnamefont {A.}~\bibnamefont {Oshiyama}},\ }\href@noop {} {\bibfield  {journal} {\bibinfo  {journal} {Physical Review B}\ }\textbf {\bibinfo {volume} {90}},\ \bibinfo {pages} {155451} (\bibinfo {year} {2014})}\BibitemShut {NoStop}%
\bibitem [{\citenamefont {Zhang}\ \emph {et~al.}(2024{\natexlab{a}})\citenamefont {Zhang}, \citenamefont {Wang}, \citenamefont {Liu}, \citenamefont {Fan}, \citenamefont {Cao},\ and\ \citenamefont {Xiao}}]{zhang2024polarization}%
  \BibitemOpen
  \bibfield  {author} {\bibinfo {author} {\bibfnamefont {X.-W.}\ \bibnamefont {Zhang}}, \bibinfo {author} {\bibfnamefont {C.}~\bibnamefont {Wang}}, \bibinfo {author} {\bibfnamefont {X.}~\bibnamefont {Liu}}, \bibinfo {author} {\bibfnamefont {Y.}~\bibnamefont {Fan}}, \bibinfo {author} {\bibfnamefont {T.}~\bibnamefont {Cao}},\ and\ \bibinfo {author} {\bibfnamefont {D.}~\bibnamefont {Xiao}},\ }\href@noop {} {\bibfield  {journal} {\bibinfo  {journal} {Nature Communications}\ }\textbf {\bibinfo {volume} {15}},\ \bibinfo {pages} {4223} (\bibinfo {year} {2024}{\natexlab{a}})}\BibitemShut {NoStop}%
\bibitem [{\citenamefont {Mao}\ \emph {et~al.}(2024)\citenamefont {Mao}, \citenamefont {Xu}, \citenamefont {Li}, \citenamefont {Bao}, \citenamefont {Liu}, \citenamefont {Xu}, \citenamefont {Felser}, \citenamefont {Fu},\ and\ \citenamefont {Zhang}}]{mao2024transfer}%
  \BibitemOpen
  \bibfield  {author} {\bibinfo {author} {\bibfnamefont {N.}~\bibnamefont {Mao}}, \bibinfo {author} {\bibfnamefont {C.}~\bibnamefont {Xu}}, \bibinfo {author} {\bibfnamefont {J.}~\bibnamefont {Li}}, \bibinfo {author} {\bibfnamefont {T.}~\bibnamefont {Bao}}, \bibinfo {author} {\bibfnamefont {P.}~\bibnamefont {Liu}}, \bibinfo {author} {\bibfnamefont {Y.}~\bibnamefont {Xu}}, \bibinfo {author} {\bibfnamefont {C.}~\bibnamefont {Felser}}, \bibinfo {author} {\bibfnamefont {L.}~\bibnamefont {Fu}},\ and\ \bibinfo {author} {\bibfnamefont {Y.}~\bibnamefont {Zhang}},\ }\href@noop {} {\bibfield  {journal} {\bibinfo  {journal} {Communications Physics}\ }\textbf {\bibinfo {volume} {7}},\ \bibinfo {pages} {262} (\bibinfo {year} {2024})}\BibitemShut {NoStop}%
\bibitem [{\citenamefont {Zhu}\ \emph {et~al.}(2026)\citenamefont {Zhu}, \citenamefont {Bennett}, \citenamefont {Larson}, \citenamefont {Ezzi}, \citenamefont {Manousakis},\ and\ \citenamefont {Kaxiras}}]{zhu2026twisted}%
  \BibitemOpen
  \bibfield  {author} {\bibinfo {author} {\bibfnamefont {A.}~\bibnamefont {Zhu}}, \bibinfo {author} {\bibfnamefont {D.}~\bibnamefont {Bennett}}, \bibinfo {author} {\bibfnamefont {D.~T.}\ \bibnamefont {Larson}}, \bibinfo {author} {\bibfnamefont {M.~M.~A.}\ \bibnamefont {Ezzi}}, \bibinfo {author} {\bibfnamefont {E.}~\bibnamefont {Manousakis}},\ and\ \bibinfo {author} {\bibfnamefont {E.}~\bibnamefont {Kaxiras}},\ }\href@noop {} {\bibfield  {journal} {\bibinfo  {journal} {Physical Review B}\ }\textbf {\bibinfo {volume} {113}},\ \bibinfo {pages} {205125} (\bibinfo {year} {2026})}\BibitemShut {NoStop}%
\bibitem [{\citenamefont {Lopes~dos Santos}\ \emph {et~al.}(2007)\citenamefont {Lopes~dos Santos}, \citenamefont {Peres},\ and\ \citenamefont {Castro~Neto}}]{lopes2007graphene}%
  \BibitemOpen
  \bibfield  {author} {\bibinfo {author} {\bibfnamefont {J.}~\bibnamefont {Lopes~dos Santos}}, \bibinfo {author} {\bibfnamefont {N.}~\bibnamefont {Peres}},\ and\ \bibinfo {author} {\bibfnamefont {A.}~\bibnamefont {Castro~Neto}},\ }\href@noop {} {\bibfield  {journal} {\bibinfo  {journal} {Physical review letters}\ }\textbf {\bibinfo {volume} {99}},\ \bibinfo {pages} {256802} (\bibinfo {year} {2007})}\BibitemShut {NoStop}%
\bibitem [{\citenamefont {Bistritzer}\ and\ \citenamefont {MacDonald}(2011)}]{bistritzer2011moire}%
  \BibitemOpen
  \bibfield  {author} {\bibinfo {author} {\bibfnamefont {R.}~\bibnamefont {Bistritzer}}\ and\ \bibinfo {author} {\bibfnamefont {A.~H.}\ \bibnamefont {MacDonald}},\ }\href@noop {} {\bibfield  {journal} {\bibinfo  {journal} {Proceedings of the National Academy of Sciences}\ }\textbf {\bibinfo {volume} {108}},\ \bibinfo {pages} {12233} (\bibinfo {year} {2011})}\BibitemShut {NoStop}%
\bibitem [{\citenamefont {De~Gail}\ \emph {et~al.}(2011)\citenamefont {De~Gail}, \citenamefont {Goerbig}, \citenamefont {Guinea}, \citenamefont {Montambaux},\ and\ \citenamefont {Castro~Neto}}]{de2011topologically}%
  \BibitemOpen
  \bibfield  {author} {\bibinfo {author} {\bibfnamefont {R.}~\bibnamefont {De~Gail}}, \bibinfo {author} {\bibfnamefont {M.}~\bibnamefont {Goerbig}}, \bibinfo {author} {\bibfnamefont {F.}~\bibnamefont {Guinea}}, \bibinfo {author} {\bibfnamefont {G.}~\bibnamefont {Montambaux}},\ and\ \bibinfo {author} {\bibfnamefont {A.}~\bibnamefont {Castro~Neto}},\ }\href@noop {} {\bibfield  {journal} {\bibinfo  {journal} {Physical Review B—Condensed Matter and Materials Physics}\ }\textbf {\bibinfo {volume} {84}},\ \bibinfo {pages} {045436} (\bibinfo {year} {2011})}\BibitemShut {NoStop}%
\bibitem [{\citenamefont {Moon}\ and\ \citenamefont {Koshino}(2012)}]{moon2012energy}%
  \BibitemOpen
  \bibfield  {author} {\bibinfo {author} {\bibfnamefont {P.}~\bibnamefont {Moon}}\ and\ \bibinfo {author} {\bibfnamefont {M.}~\bibnamefont {Koshino}},\ }\href@noop {} {\bibfield  {journal} {\bibinfo  {journal} {Physical Review B—Condensed Matter and Materials Physics}\ }\textbf {\bibinfo {volume} {85}},\ \bibinfo {pages} {195458} (\bibinfo {year} {2012})}\BibitemShut {NoStop}%
\bibitem [{\citenamefont {Jung}\ \emph {et~al.}(2014)\citenamefont {Jung}, \citenamefont {Raoux}, \citenamefont {Qiao},\ and\ \citenamefont {MacDonald}}]{jung2014ab}%
  \BibitemOpen
  \bibfield  {author} {\bibinfo {author} {\bibfnamefont {J.}~\bibnamefont {Jung}}, \bibinfo {author} {\bibfnamefont {A.}~\bibnamefont {Raoux}}, \bibinfo {author} {\bibfnamefont {Z.}~\bibnamefont {Qiao}},\ and\ \bibinfo {author} {\bibfnamefont {A.~H.}\ \bibnamefont {MacDonald}},\ }\href@noop {} {\bibfield  {journal} {\bibinfo  {journal} {Physical Review B}\ }\textbf {\bibinfo {volume} {89}},\ \bibinfo {pages} {205414} (\bibinfo {year} {2014})}\BibitemShut {NoStop}%
\bibitem [{\citenamefont {Wu}\ \emph {et~al.}(2019)\citenamefont {Wu}, \citenamefont {Lovorn}, \citenamefont {Tutuc}, \citenamefont {Martin},\ and\ \citenamefont {MacDonald}}]{wu2019topological}%
  \BibitemOpen
  \bibfield  {author} {\bibinfo {author} {\bibfnamefont {F.}~\bibnamefont {Wu}}, \bibinfo {author} {\bibfnamefont {T.}~\bibnamefont {Lovorn}}, \bibinfo {author} {\bibfnamefont {E.}~\bibnamefont {Tutuc}}, \bibinfo {author} {\bibfnamefont {I.}~\bibnamefont {Martin}},\ and\ \bibinfo {author} {\bibfnamefont {A.~H.}\ \bibnamefont {MacDonald}},\ }\href@noop {} {\bibfield  {journal} {\bibinfo  {journal} {Physical review letters}\ }\textbf {\bibinfo {volume} {122}},\ \bibinfo {pages} {086402} (\bibinfo {year} {2019})}\BibitemShut {NoStop}%
\bibitem [{\citenamefont {Devakul}\ \emph {et~al.}(2021)\citenamefont {Devakul}, \citenamefont {Cr{\'e}pel}, \citenamefont {Zhang},\ and\ \citenamefont {Fu}}]{devakul2021magic}%
  \BibitemOpen
  \bibfield  {author} {\bibinfo {author} {\bibfnamefont {T.}~\bibnamefont {Devakul}}, \bibinfo {author} {\bibfnamefont {V.}~\bibnamefont {Cr{\'e}pel}}, \bibinfo {author} {\bibfnamefont {Y.}~\bibnamefont {Zhang}},\ and\ \bibinfo {author} {\bibfnamefont {L.}~\bibnamefont {Fu}},\ }\href@noop {} {\bibfield  {journal} {\bibinfo  {journal} {Nature communications}\ }\textbf {\bibinfo {volume} {12}},\ \bibinfo {pages} {6730} (\bibinfo {year} {2021})}\BibitemShut {NoStop}%
\bibitem [{\citenamefont {Jia}\ \emph {et~al.}(2024)\citenamefont {Jia}, \citenamefont {Yu}, \citenamefont {Liu}, \citenamefont {Herzog-Arbeitman}, \citenamefont {Qi}, \citenamefont {Pi}, \citenamefont {Regnault}, \citenamefont {Weng}, \citenamefont {Bernevig},\ and\ \citenamefont {Wu}}]{jia2024moire}%
  \BibitemOpen
  \bibfield  {author} {\bibinfo {author} {\bibfnamefont {Y.}~\bibnamefont {Jia}}, \bibinfo {author} {\bibfnamefont {J.}~\bibnamefont {Yu}}, \bibinfo {author} {\bibfnamefont {J.}~\bibnamefont {Liu}}, \bibinfo {author} {\bibfnamefont {J.}~\bibnamefont {Herzog-Arbeitman}}, \bibinfo {author} {\bibfnamefont {Z.}~\bibnamefont {Qi}}, \bibinfo {author} {\bibfnamefont {H.}~\bibnamefont {Pi}}, \bibinfo {author} {\bibfnamefont {N.}~\bibnamefont {Regnault}}, \bibinfo {author} {\bibfnamefont {H.}~\bibnamefont {Weng}}, \bibinfo {author} {\bibfnamefont {B.~A.}\ \bibnamefont {Bernevig}},\ and\ \bibinfo {author} {\bibfnamefont {Q.}~\bibnamefont {Wu}},\ }\href@noop {} {\bibfield  {journal} {\bibinfo  {journal} {Physical Review B}\ }\textbf {\bibinfo {volume} {109}},\ \bibinfo {pages} {205121} (\bibinfo {year} {2024})}\BibitemShut {NoStop}%
\bibitem [{\citenamefont {Cr{\'e}pel}\ and\ \citenamefont {Fu}(2023)}]{crepel2023anomalous}%
  \BibitemOpen
  \bibfield  {author} {\bibinfo {author} {\bibfnamefont {V.}~\bibnamefont {Cr{\'e}pel}}\ and\ \bibinfo {author} {\bibfnamefont {L.}~\bibnamefont {Fu}},\ }\href@noop {} {\bibfield  {journal} {\bibinfo  {journal} {Physical Review B}\ }\textbf {\bibinfo {volume} {107}},\ \bibinfo {pages} {L201109} (\bibinfo {year} {2023})}\BibitemShut {NoStop}%
\bibitem [{\citenamefont {Cai}\ \emph {et~al.}(2023)\citenamefont {Cai}, \citenamefont {Anderson}, \citenamefont {Wang}, \citenamefont {Zhang}, \citenamefont {Liu}, \citenamefont {Holtzmann}, \citenamefont {Zhang}, \citenamefont {Fan}, \citenamefont {Taniguchi}, \citenamefont {Watanabe} \emph {et~al.}}]{cai2023signatures}%
  \BibitemOpen
  \bibfield  {author} {\bibinfo {author} {\bibfnamefont {J.}~\bibnamefont {Cai}}, \bibinfo {author} {\bibfnamefont {E.}~\bibnamefont {Anderson}}, \bibinfo {author} {\bibfnamefont {C.}~\bibnamefont {Wang}}, \bibinfo {author} {\bibfnamefont {X.}~\bibnamefont {Zhang}}, \bibinfo {author} {\bibfnamefont {X.}~\bibnamefont {Liu}}, \bibinfo {author} {\bibfnamefont {W.}~\bibnamefont {Holtzmann}}, \bibinfo {author} {\bibfnamefont {Y.}~\bibnamefont {Zhang}}, \bibinfo {author} {\bibfnamefont {F.}~\bibnamefont {Fan}}, \bibinfo {author} {\bibfnamefont {T.}~\bibnamefont {Taniguchi}}, \bibinfo {author} {\bibfnamefont {K.}~\bibnamefont {Watanabe}}, \emph {et~al.},\ }\href@noop {} {\bibfield  {journal} {\bibinfo  {journal} {Nature}\ }\textbf {\bibinfo {volume} {622}},\ \bibinfo {pages} {63} (\bibinfo {year} {2023})}\BibitemShut {NoStop}%
\bibitem [{\citenamefont {Zeng}\ \emph {et~al.}(2023)\citenamefont {Zeng}, \citenamefont {Xia}, \citenamefont {Kang}, \citenamefont {Zhu}, \citenamefont {Kn{\"u}ppel}, \citenamefont {Vaswani}, \citenamefont {Watanabe}, \citenamefont {Taniguchi}, \citenamefont {Mak},\ and\ \citenamefont {Shan}}]{zeng2023thermodynamic}%
  \BibitemOpen
  \bibfield  {author} {\bibinfo {author} {\bibfnamefont {Y.}~\bibnamefont {Zeng}}, \bibinfo {author} {\bibfnamefont {Z.}~\bibnamefont {Xia}}, \bibinfo {author} {\bibfnamefont {K.}~\bibnamefont {Kang}}, \bibinfo {author} {\bibfnamefont {J.}~\bibnamefont {Zhu}}, \bibinfo {author} {\bibfnamefont {P.}~\bibnamefont {Kn{\"u}ppel}}, \bibinfo {author} {\bibfnamefont {C.}~\bibnamefont {Vaswani}}, \bibinfo {author} {\bibfnamefont {K.}~\bibnamefont {Watanabe}}, \bibinfo {author} {\bibfnamefont {T.}~\bibnamefont {Taniguchi}}, \bibinfo {author} {\bibfnamefont {K.~F.}\ \bibnamefont {Mak}},\ and\ \bibinfo {author} {\bibfnamefont {J.}~\bibnamefont {Shan}},\ }\href@noop {} {\bibfield  {journal} {\bibinfo  {journal} {Nature}\ }\textbf {\bibinfo {volume} {622}},\ \bibinfo {pages} {69} (\bibinfo {year} {2023})}\BibitemShut {NoStop}%
\bibitem [{\citenamefont {Fischer}\ \emph {et~al.}(2024)\citenamefont {Fischer}, \citenamefont {Klebl}, \citenamefont {Cr\'epel}, \citenamefont {Ryee}, \citenamefont {Rubio}, \citenamefont {Xian}, \citenamefont {Wehling}, \citenamefont {Georges}, \citenamefont {Kennes},\ and\ \citenamefont {Millis}}]{fischer2024theory}%
  \BibitemOpen
  \bibfield  {author} {\bibinfo {author} {\bibfnamefont {A.}~\bibnamefont {Fischer}}, \bibinfo {author} {\bibfnamefont {L.}~\bibnamefont {Klebl}}, \bibinfo {author} {\bibfnamefont {V.}~\bibnamefont {Cr\'epel}}, \bibinfo {author} {\bibfnamefont {S.}~\bibnamefont {Ryee}}, \bibinfo {author} {\bibfnamefont {A.}~\bibnamefont {Rubio}}, \bibinfo {author} {\bibfnamefont {L.}~\bibnamefont {Xian}}, \bibinfo {author} {\bibfnamefont {T.~O.}\ \bibnamefont {Wehling}}, \bibinfo {author} {\bibfnamefont {A.}~\bibnamefont {Georges}}, \bibinfo {author} {\bibfnamefont {D.~M.}\ \bibnamefont {Kennes}},\ and\ \bibinfo {author} {\bibfnamefont {A.~J.}\ \bibnamefont {Millis}},\ }\href@noop {} {\bibfield  {journal} {\bibinfo  {journal} {arXiv preprint arXiv:2412.14296}\ } (\bibinfo {year} {2024})}\BibitemShut {NoStop}%
\bibitem [{\citenamefont {Mu{\~n}oz-Segovia}\ \emph {et~al.}(2025)\citenamefont {Mu{\~n}oz-Segovia}, \citenamefont {Cr{\'e}pel}, \citenamefont {Queiroz},\ and\ \citenamefont {Millis}}]{munoz2025twist}%
  \BibitemOpen
  \bibfield  {author} {\bibinfo {author} {\bibfnamefont {D.}~\bibnamefont {Mu{\~n}oz-Segovia}}, \bibinfo {author} {\bibfnamefont {V.}~\bibnamefont {Cr{\'e}pel}}, \bibinfo {author} {\bibfnamefont {R.}~\bibnamefont {Queiroz}},\ and\ \bibinfo {author} {\bibfnamefont {A.~J.}\ \bibnamefont {Millis}},\ }\href@noop {} {\bibfield  {journal} {\bibinfo  {journal} {Physical Review B}\ }\textbf {\bibinfo {volume} {112}},\ \bibinfo {pages} {085111} (\bibinfo {year} {2025})}\BibitemShut {NoStop}%
\bibitem [{\citenamefont {Guo}\ \emph {et~al.}(2025)\citenamefont {Guo}, \citenamefont {Pack}, \citenamefont {Swann}, \citenamefont {Holtzman}, \citenamefont {Cothrine}, \citenamefont {Watanabe}, \citenamefont {Taniguchi}, \citenamefont {Mandrus}, \citenamefont {Barmak}, \citenamefont {Hone} \emph {et~al.}}]{guo2025superconductivity}%
  \BibitemOpen
  \bibfield  {author} {\bibinfo {author} {\bibfnamefont {Y.}~\bibnamefont {Guo}}, \bibinfo {author} {\bibfnamefont {J.}~\bibnamefont {Pack}}, \bibinfo {author} {\bibfnamefont {J.}~\bibnamefont {Swann}}, \bibinfo {author} {\bibfnamefont {L.}~\bibnamefont {Holtzman}}, \bibinfo {author} {\bibfnamefont {M.}~\bibnamefont {Cothrine}}, \bibinfo {author} {\bibfnamefont {K.}~\bibnamefont {Watanabe}}, \bibinfo {author} {\bibfnamefont {T.}~\bibnamefont {Taniguchi}}, \bibinfo {author} {\bibfnamefont {D.~G.}\ \bibnamefont {Mandrus}}, \bibinfo {author} {\bibfnamefont {K.}~\bibnamefont {Barmak}}, \bibinfo {author} {\bibfnamefont {J.}~\bibnamefont {Hone}}, \emph {et~al.},\ }\href@noop {} {\bibfield  {journal} {\bibinfo  {journal} {Nature}\ }\textbf {\bibinfo {volume} {637}},\ \bibinfo {pages} {839} (\bibinfo {year} {2025})}\BibitemShut {NoStop}%
\bibitem [{\citenamefont {Xia}\ \emph {et~al.}(2025)\citenamefont {Xia}, \citenamefont {Han}, \citenamefont {Watanabe}, \citenamefont {Taniguchi}, \citenamefont {Shan},\ and\ \citenamefont {Mak}}]{xia2025superconductivity}%
  \BibitemOpen
  \bibfield  {author} {\bibinfo {author} {\bibfnamefont {Y.}~\bibnamefont {Xia}}, \bibinfo {author} {\bibfnamefont {Z.}~\bibnamefont {Han}}, \bibinfo {author} {\bibfnamefont {K.}~\bibnamefont {Watanabe}}, \bibinfo {author} {\bibfnamefont {T.}~\bibnamefont {Taniguchi}}, \bibinfo {author} {\bibfnamefont {J.}~\bibnamefont {Shan}},\ and\ \bibinfo {author} {\bibfnamefont {K.~F.}\ \bibnamefont {Mak}},\ }\href@noop {} {\bibfield  {journal} {\bibinfo  {journal} {Nature}\ }\textbf {\bibinfo {volume} {637}},\ \bibinfo {pages} {833} (\bibinfo {year} {2025})}\BibitemShut {NoStop}%
\bibitem [{\citenamefont {Haastrup}\ \emph {et~al.}(2018)\citenamefont {Haastrup}, \citenamefont {Strange}, \citenamefont {Pandey}, \citenamefont {Deilmann}, \citenamefont {Schmidt}, \citenamefont {Hinsche}, \citenamefont {Gjerding}, \citenamefont {Torelli}, \citenamefont {Larsen}, \citenamefont {Riis-Jensen} \emph {et~al.}}]{haastrup2018computational}%
  \BibitemOpen
  \bibfield  {author} {\bibinfo {author} {\bibfnamefont {S.}~\bibnamefont {Haastrup}}, \bibinfo {author} {\bibfnamefont {M.}~\bibnamefont {Strange}}, \bibinfo {author} {\bibfnamefont {M.}~\bibnamefont {Pandey}}, \bibinfo {author} {\bibfnamefont {T.}~\bibnamefont {Deilmann}}, \bibinfo {author} {\bibfnamefont {P.~S.}\ \bibnamefont {Schmidt}}, \bibinfo {author} {\bibfnamefont {N.~F.}\ \bibnamefont {Hinsche}}, \bibinfo {author} {\bibfnamefont {M.~N.}\ \bibnamefont {Gjerding}}, \bibinfo {author} {\bibfnamefont {D.}~\bibnamefont {Torelli}}, \bibinfo {author} {\bibfnamefont {P.~M.}\ \bibnamefont {Larsen}}, \bibinfo {author} {\bibfnamefont {A.~C.}\ \bibnamefont {Riis-Jensen}}, \emph {et~al.},\ }\href@noop {} {\bibfield  {journal} {\bibinfo  {journal} {2D Materials}\ }\textbf {\bibinfo {volume} {5}},\ \bibinfo {pages} {042002} (\bibinfo {year} {2018})}\BibitemShut {NoStop}%
\bibitem [{\citenamefont {Gjerding}\ \emph {et~al.}(2021)\citenamefont {Gjerding}, \citenamefont {Taghizadeh}, \citenamefont {Rasmussen}, \citenamefont {Ali}, \citenamefont {Bertoldo}, \citenamefont {Deilmann}, \citenamefont {Kn{\o}sgaard}, \citenamefont {Kruse}, \citenamefont {Larsen}, \citenamefont {Manti} \emph {et~al.}}]{gjerding2021recent}%
  \BibitemOpen
  \bibfield  {author} {\bibinfo {author} {\bibfnamefont {M.~N.}\ \bibnamefont {Gjerding}}, \bibinfo {author} {\bibfnamefont {A.}~\bibnamefont {Taghizadeh}}, \bibinfo {author} {\bibfnamefont {A.}~\bibnamefont {Rasmussen}}, \bibinfo {author} {\bibfnamefont {S.}~\bibnamefont {Ali}}, \bibinfo {author} {\bibfnamefont {F.}~\bibnamefont {Bertoldo}}, \bibinfo {author} {\bibfnamefont {T.}~\bibnamefont {Deilmann}}, \bibinfo {author} {\bibfnamefont {N.~R.}\ \bibnamefont {Kn{\o}sgaard}}, \bibinfo {author} {\bibfnamefont {M.}~\bibnamefont {Kruse}}, \bibinfo {author} {\bibfnamefont {A.~H.}\ \bibnamefont {Larsen}}, \bibinfo {author} {\bibfnamefont {S.}~\bibnamefont {Manti}}, \emph {et~al.},\ }\href@noop {} {\bibfield  {journal} {\bibinfo  {journal} {2D Materials}\ }\textbf {\bibinfo {volume} {8}},\ \bibinfo {pages} {044002} (\bibinfo {year} {2021})}\BibitemShut {NoStop}%
\bibitem [{\citenamefont {Rice}\ and\ \citenamefont {Scott}(1975)}]{rice1975new}%
  \BibitemOpen
  \bibfield  {author} {\bibinfo {author} {\bibfnamefont {T.}~\bibnamefont {Rice}}\ and\ \bibinfo {author} {\bibfnamefont {G.}~\bibnamefont {Scott}},\ }\href@noop {} {\bibfield  {journal} {\bibinfo  {journal} {Physical Review Letters}\ }\textbf {\bibinfo {volume} {35}},\ \bibinfo {pages} {120} (\bibinfo {year} {1975})}\BibitemShut {NoStop}%
\bibitem [{\citenamefont {Ugeda}\ \emph {et~al.}(2016)\citenamefont {Ugeda}, \citenamefont {Bradley}, \citenamefont {Zhang}, \citenamefont {Onishi}, \citenamefont {Chen}, \citenamefont {Ruan}, \citenamefont {Ojeda-Aristizabal}, \citenamefont {Ryu}, \citenamefont {Edmonds}, \citenamefont {Tsai}, \citenamefont {Riss}, \citenamefont {Mo}, \citenamefont {Lee}, \citenamefont {Zettl}, \citenamefont {Hussain}, \citenamefont {Shen},\ and\ \citenamefont {Crommie}}]{ugeda2016characterization}%
  \BibitemOpen
  \bibfield  {author} {\bibinfo {author} {\bibfnamefont {M.~M.}\ \bibnamefont {Ugeda}}, \bibinfo {author} {\bibfnamefont {A.~J.}\ \bibnamefont {Bradley}}, \bibinfo {author} {\bibfnamefont {Y.}~\bibnamefont {Zhang}}, \bibinfo {author} {\bibfnamefont {S.}~\bibnamefont {Onishi}}, \bibinfo {author} {\bibfnamefont {Y.}~\bibnamefont {Chen}}, \bibinfo {author} {\bibfnamefont {W.}~\bibnamefont {Ruan}}, \bibinfo {author} {\bibfnamefont {C.}~\bibnamefont {Ojeda-Aristizabal}}, \bibinfo {author} {\bibfnamefont {H.}~\bibnamefont {Ryu}}, \bibinfo {author} {\bibfnamefont {M.~T.}\ \bibnamefont {Edmonds}}, \bibinfo {author} {\bibfnamefont {H.-Z.}\ \bibnamefont {Tsai}}, \bibinfo {author} {\bibfnamefont {A.}~\bibnamefont {Riss}}, \bibinfo {author} {\bibfnamefont {S.-K.}\ \bibnamefont {Mo}}, \bibinfo {author} {\bibfnamefont {D.}~\bibnamefont {Lee}}, \bibinfo {author} {\bibfnamefont {A.}~\bibnamefont {Zettl}}, \bibinfo {author} {\bibfnamefont {Z.}~\bibnamefont {Hussain}}, \bibinfo {author} {\bibfnamefont {Z.-X.}\ \bibnamefont
  {Shen}},\ and\ \bibinfo {author} {\bibfnamefont {M.~F.}\ \bibnamefont {Crommie}},\ }\href {https://doi.org/10.1038/nphys3527} {\bibfield  {journal} {\bibinfo  {journal} {Nature Physics}\ }\textbf {\bibinfo {volume} {12}},\ \bibinfo {pages} {92} (\bibinfo {year} {2016})}\BibitemShut {NoStop}%
\bibitem [{\citenamefont {Xi}\ \emph {et~al.}(2016)\citenamefont {Xi}, \citenamefont {Wang}, \citenamefont {Zhao}, \citenamefont {Park}, \citenamefont {Law}, \citenamefont {Berger}, \citenamefont {Forr{\'o}}, \citenamefont {Shan},\ and\ \citenamefont {Mak}}]{xi2016ising}%
  \BibitemOpen
  \bibfield  {author} {\bibinfo {author} {\bibfnamefont {X.}~\bibnamefont {Xi}}, \bibinfo {author} {\bibfnamefont {Z.}~\bibnamefont {Wang}}, \bibinfo {author} {\bibfnamefont {W.}~\bibnamefont {Zhao}}, \bibinfo {author} {\bibfnamefont {J.-H.}\ \bibnamefont {Park}}, \bibinfo {author} {\bibfnamefont {K.~T.}\ \bibnamefont {Law}}, \bibinfo {author} {\bibfnamefont {H.}~\bibnamefont {Berger}}, \bibinfo {author} {\bibfnamefont {L.}~\bibnamefont {Forr{\'o}}}, \bibinfo {author} {\bibfnamefont {J.}~\bibnamefont {Shan}},\ and\ \bibinfo {author} {\bibfnamefont {K.~F.}\ \bibnamefont {Mak}},\ }\href {https://doi.org/10.1038/nphys3538} {\bibfield  {journal} {\bibinfo  {journal} {Nature Physics}\ }\textbf {\bibinfo {volume} {12}},\ \bibinfo {pages} {139} (\bibinfo {year} {2016})}\BibitemShut {NoStop}%
\bibitem [{\citenamefont {Nakata}\ \emph {et~al.}(2018)\citenamefont {Nakata}, \citenamefont {Sugawara}, \citenamefont {Ichinokura}, \citenamefont {Okada}, \citenamefont {Hitosugi}, \citenamefont {Koretsune}, \citenamefont {Ueno}, \citenamefont {Hasegawa}, \citenamefont {Takahashi},\ and\ \citenamefont {Sato}}]{nakata2018anisotropic}%
  \BibitemOpen
  \bibfield  {author} {\bibinfo {author} {\bibfnamefont {Y.}~\bibnamefont {Nakata}}, \bibinfo {author} {\bibfnamefont {K.}~\bibnamefont {Sugawara}}, \bibinfo {author} {\bibfnamefont {S.}~\bibnamefont {Ichinokura}}, \bibinfo {author} {\bibfnamefont {Y.}~\bibnamefont {Okada}}, \bibinfo {author} {\bibfnamefont {T.}~\bibnamefont {Hitosugi}}, \bibinfo {author} {\bibfnamefont {T.}~\bibnamefont {Koretsune}}, \bibinfo {author} {\bibfnamefont {K.}~\bibnamefont {Ueno}}, \bibinfo {author} {\bibfnamefont {S.}~\bibnamefont {Hasegawa}}, \bibinfo {author} {\bibfnamefont {T.}~\bibnamefont {Takahashi}},\ and\ \bibinfo {author} {\bibfnamefont {T.}~\bibnamefont {Sato}},\ }\href@noop {} {\bibfield  {journal} {\bibinfo  {journal} {npj 2D Materials and Applications}\ }\textbf {\bibinfo {volume} {2}},\ \bibinfo {pages} {12} (\bibinfo {year} {2018})}\BibitemShut {NoStop}%
\bibitem [{\citenamefont {Cheung}\ \emph {et~al.}(2024)\citenamefont {Cheung}, \citenamefont {Goodwin}, \citenamefont {Han}, \citenamefont {Lu}, \citenamefont {Mostofi},\ and\ \citenamefont {Lischner}}]{cheung2024coexisting}%
  \BibitemOpen
  \bibfield  {author} {\bibinfo {author} {\bibfnamefont {C.~T.}\ \bibnamefont {Cheung}}, \bibinfo {author} {\bibfnamefont {Z.~A.}\ \bibnamefont {Goodwin}}, \bibinfo {author} {\bibfnamefont {Y.}~\bibnamefont {Han}}, \bibinfo {author} {\bibfnamefont {J.}~\bibnamefont {Lu}}, \bibinfo {author} {\bibfnamefont {A.~A.}\ \bibnamefont {Mostofi}},\ and\ \bibinfo {author} {\bibfnamefont {J.}~\bibnamefont {Lischner}},\ }\href@noop {} {\bibfield  {journal} {\bibinfo  {journal} {Nano Letters}\ }\textbf {\bibinfo {volume} {24}},\ \bibinfo {pages} {12088} (\bibinfo {year} {2024})}\BibitemShut {NoStop}%
\bibitem [{\citenamefont {Slater}\ and\ \citenamefont {Koster}(1954)}]{slater1954simplified}%
  \BibitemOpen
  \bibfield  {author} {\bibinfo {author} {\bibfnamefont {J.~C.}\ \bibnamefont {Slater}}\ and\ \bibinfo {author} {\bibfnamefont {G.~F.}\ \bibnamefont {Koster}},\ }\href@noop {} {\bibfield  {journal} {\bibinfo  {journal} {Physical review}\ }\textbf {\bibinfo {volume} {94}},\ \bibinfo {pages} {1498} (\bibinfo {year} {1954})}\BibitemShut {NoStop}%
\bibitem [{\citenamefont {Watson}\ \emph {et~al.}(2023)\citenamefont {Watson}, \citenamefont {Kong}, \citenamefont {MacDonald},\ and\ \citenamefont {Luskin}}]{watson2023bistritzer}%
  \BibitemOpen
  \bibfield  {author} {\bibinfo {author} {\bibfnamefont {A.~B.}\ \bibnamefont {Watson}}, \bibinfo {author} {\bibfnamefont {T.}~\bibnamefont {Kong}}, \bibinfo {author} {\bibfnamefont {A.~H.}\ \bibnamefont {MacDonald}},\ and\ \bibinfo {author} {\bibfnamefont {M.}~\bibnamefont {Luskin}},\ }\href@noop {} {\bibfield  {journal} {\bibinfo  {journal} {Journal of Mathematical Physics}\ }\textbf {\bibinfo {volume} {64}} (\bibinfo {year} {2023})}\BibitemShut {NoStop}%
\bibitem [{\citenamefont {Quinn}\ \emph {et~al.}(2025)\citenamefont {Quinn}, \citenamefont {Kong}, \citenamefont {Luskin},\ and\ \citenamefont {Watson}}]{quinn2025higher}%
  \BibitemOpen
  \bibfield  {author} {\bibinfo {author} {\bibfnamefont {S.}~\bibnamefont {Quinn}}, \bibinfo {author} {\bibfnamefont {T.}~\bibnamefont {Kong}}, \bibinfo {author} {\bibfnamefont {M.}~\bibnamefont {Luskin}},\ and\ \bibinfo {author} {\bibfnamefont {A.~B.}\ \bibnamefont {Watson}},\ }\href@noop {} {\bibfield  {journal} {\bibinfo  {journal} {Journal of Mathematical Physics}\ }\textbf {\bibinfo {volume} {66}} (\bibinfo {year} {2025})}\BibitemShut {NoStop}%
\bibitem [{\citenamefont {Kang}\ and\ \citenamefont {Vafek}(2023)}]{kang2023pseudomagnetic}%
  \BibitemOpen
  \bibfield  {author} {\bibinfo {author} {\bibfnamefont {J.}~\bibnamefont {Kang}}\ and\ \bibinfo {author} {\bibfnamefont {O.}~\bibnamefont {Vafek}},\ }\href@noop {} {\bibfield  {journal} {\bibinfo  {journal} {Phys. Rev. B}\ }\textbf {\bibinfo {volume} {107}},\ \bibinfo {pages} {075408} (\bibinfo {year} {2023})}\BibitemShut {NoStop}%
\bibitem [{\citenamefont {Vafek}\ and\ \citenamefont {Kang}(2023)}]{vafek2023continuum}%
  \BibitemOpen
  \bibfield  {author} {\bibinfo {author} {\bibfnamefont {O.}~\bibnamefont {Vafek}}\ and\ \bibinfo {author} {\bibfnamefont {J.}~\bibnamefont {Kang}},\ }\href@noop {} {\bibfield  {journal} {\bibinfo  {journal} {Phys. Rev. B}\ }\textbf {\bibinfo {volume} {107}},\ \bibinfo {pages} {075123} (\bibinfo {year} {2023})}\BibitemShut {NoStop}%
\bibitem [{\citenamefont {Zhang}\ \emph {et~al.}(2021)\citenamefont {Zhang}, \citenamefont {Liu},\ and\ \citenamefont {Fu}}]{zhang2021electronic}%
  \BibitemOpen
  \bibfield  {author} {\bibinfo {author} {\bibfnamefont {Y.}~\bibnamefont {Zhang}}, \bibinfo {author} {\bibfnamefont {T.}~\bibnamefont {Liu}},\ and\ \bibinfo {author} {\bibfnamefont {L.}~\bibnamefont {Fu}},\ }\href@noop {} {\bibfield  {journal} {\bibinfo  {journal} {Physical Review B}\ }\textbf {\bibinfo {volume} {103}},\ \bibinfo {pages} {155142} (\bibinfo {year} {2021})}\BibitemShut {NoStop}%
\bibitem [{\citenamefont {Angeli}\ and\ \citenamefont {MacDonald}(2021)}]{angeli2021gamma}%
  \BibitemOpen
  \bibfield  {author} {\bibinfo {author} {\bibfnamefont {M.}~\bibnamefont {Angeli}}\ and\ \bibinfo {author} {\bibfnamefont {A.~H.}\ \bibnamefont {MacDonald}},\ }\href@noop {} {\bibfield  {journal} {\bibinfo  {journal} {Proceedings of the National Academy of Sciences}\ }\textbf {\bibinfo {volume} {118}},\ \bibinfo {pages} {e2021826118} (\bibinfo {year} {2021})}\BibitemShut {NoStop}%
\bibitem [{\citenamefont {Lei}\ \emph {et~al.}(2025)\citenamefont {Lei}, \citenamefont {Mahon},\ and\ \citenamefont {MacDonald}}]{5zt2-scbg}%
  \BibitemOpen
  \bibfield  {author} {\bibinfo {author} {\bibfnamefont {C.}~\bibnamefont {Lei}}, \bibinfo {author} {\bibfnamefont {P.~T.}\ \bibnamefont {Mahon}},\ and\ \bibinfo {author} {\bibfnamefont {A.~H.}\ \bibnamefont {MacDonald}},\ }\href@noop {} {\bibfield  {journal} {\bibinfo  {journal} {Physical Review Letters}\ }\textbf {\bibinfo {volume} {135}},\ \bibinfo {pages} {196402} (\bibinfo {year} {2025})}\BibitemShut {NoStop}%
\bibitem [{\citenamefont {Cr{\'e}pel}\ \emph {et~al.}(2024)\citenamefont {Cr{\'e}pel}, \citenamefont {Regnault},\ and\ \citenamefont {Queiroz}}]{crepel2024chiral}%
  \BibitemOpen
  \bibfield  {author} {\bibinfo {author} {\bibfnamefont {V.}~\bibnamefont {Cr{\'e}pel}}, \bibinfo {author} {\bibfnamefont {N.}~\bibnamefont {Regnault}},\ and\ \bibinfo {author} {\bibfnamefont {R.}~\bibnamefont {Queiroz}},\ }\href@noop {} {\bibfield  {journal} {\bibinfo  {journal} {Communications Physics}\ }\textbf {\bibinfo {volume} {7}},\ \bibinfo {pages} {146} (\bibinfo {year} {2024})}\BibitemShut {NoStop}%
\bibitem [{\citenamefont {Zhang}\ \emph {et~al.}(2025)\citenamefont {Zhang}, \citenamefont {Yang}, \citenamefont {Wang}, \citenamefont {Liu}, \citenamefont {Cao},\ and\ \citenamefont {Xiao}}]{zhang2025twist}%
  \BibitemOpen
  \bibfield  {author} {\bibinfo {author} {\bibfnamefont {X.-W.}\ \bibnamefont {Zhang}}, \bibinfo {author} {\bibfnamefont {K.}~\bibnamefont {Yang}}, \bibinfo {author} {\bibfnamefont {C.}~\bibnamefont {Wang}}, \bibinfo {author} {\bibfnamefont {X.}~\bibnamefont {Liu}}, \bibinfo {author} {\bibfnamefont {T.}~\bibnamefont {Cao}},\ and\ \bibinfo {author} {\bibfnamefont {D.}~\bibnamefont {Xiao}},\ }\href@noop {} {\bibfield  {journal} {\bibinfo  {journal} {npj Quantum Materials}\ }\textbf {\bibinfo {volume} {10}},\ \bibinfo {pages} {110} (\bibinfo {year} {2025})}\BibitemShut {NoStop}%
\bibitem [{\citenamefont {Espitia}\ \emph {et~al.}(2025)\citenamefont {Espitia}, \citenamefont {Lee}, \citenamefont {Chiu}, \citenamefont {Ahn},\ and\ \citenamefont {Naik}}]{espitia2025quasi}%
  \BibitemOpen
  \bibfield  {author} {\bibinfo {author} {\bibfnamefont {G.}~\bibnamefont {Espitia}}, \bibinfo {author} {\bibfnamefont {S.~H.}\ \bibnamefont {Lee}}, \bibinfo {author} {\bibfnamefont {C.~K.}\ \bibnamefont {Chiu}}, \bibinfo {author} {\bibfnamefont {J.}~\bibnamefont {Ahn}},\ and\ \bibinfo {author} {\bibfnamefont {M.~H.}\ \bibnamefont {Naik}},\ }\href@noop {} {\bibfield  {journal} {\bibinfo  {journal} {arXiv preprint arXiv:2512.11137}\ } (\bibinfo {year} {2025})}\BibitemShut {NoStop}%
\bibitem [{\citenamefont {Zhang}\ \emph {et~al.}(2024{\natexlab{b}})\citenamefont {Zhang}, \citenamefont {Pi}, \citenamefont {Liu}, \citenamefont {Miao}, \citenamefont {Qi}, \citenamefont {Regnault}, \citenamefont {Weng}, \citenamefont {Dai}, \citenamefont {Bernevig}, \citenamefont {Wu} \emph {et~al.}}]{zhang2024universal}%
  \BibitemOpen
  \bibfield  {author} {\bibinfo {author} {\bibfnamefont {Y.}~\bibnamefont {Zhang}}, \bibinfo {author} {\bibfnamefont {H.}~\bibnamefont {Pi}}, \bibinfo {author} {\bibfnamefont {J.}~\bibnamefont {Liu}}, \bibinfo {author} {\bibfnamefont {W.}~\bibnamefont {Miao}}, \bibinfo {author} {\bibfnamefont {Z.}~\bibnamefont {Qi}}, \bibinfo {author} {\bibfnamefont {N.}~\bibnamefont {Regnault}}, \bibinfo {author} {\bibfnamefont {H.}~\bibnamefont {Weng}}, \bibinfo {author} {\bibfnamefont {X.}~\bibnamefont {Dai}}, \bibinfo {author} {\bibfnamefont {B.~A.}\ \bibnamefont {Bernevig}}, \bibinfo {author} {\bibfnamefont {Q.}~\bibnamefont {Wu}}, \emph {et~al.},\ }\href@noop {} {\bibfield  {journal} {\bibinfo  {journal} {arXiv preprint arXiv:2411.08108}\ } (\bibinfo {year} {2024}{\natexlab{b}})}\BibitemShut {NoStop}%
\bibitem [{\citenamefont {De~Laissardiere}\ \emph {et~al.}(2010)\citenamefont {De~Laissardiere}, \citenamefont {Mayou}, \citenamefont {Magaud} \emph {et~al.}}]{de2010localization}%
  \BibitemOpen
  \bibfield  {author} {\bibinfo {author} {\bibfnamefont {G.~T.}\ \bibnamefont {De~Laissardiere}}, \bibinfo {author} {\bibfnamefont {D.}~\bibnamefont {Mayou}}, \bibinfo {author} {\bibfnamefont {L.}~\bibnamefont {Magaud}}, \emph {et~al.},\ }\href@noop {} {\bibfield  {journal} {\bibinfo  {journal} {Nano Lett}\ }\textbf {\bibinfo {volume} {10}},\ \bibinfo {pages} {804} (\bibinfo {year} {2010})}\BibitemShut {NoStop}%
\bibitem [{\citenamefont {Pizzi}\ \emph {et~al.}(2020)\citenamefont {Pizzi}, \citenamefont {Vitale}, \citenamefont {Arita}, \citenamefont {Blügel}, \citenamefont {Freimuth}, \citenamefont {Géranton}, \citenamefont {Gibertini}, \citenamefont {Gresch}, \citenamefont {Johnson}, \citenamefont {Koretsune}, \citenamefont {Ibañez-Azpiroz}, \citenamefont {Lee}, \citenamefont {Lihm}, \citenamefont {Marchand}, \citenamefont {Marrazzo}, \citenamefont {Mokrousov}, \citenamefont {Mustafa}, \citenamefont {Nohara}, \citenamefont {Nomura}, \citenamefont {Paulatto}, \citenamefont {Poncé}, \citenamefont {Ponweiser}, \citenamefont {Qiao}, \citenamefont {Thöle}, \citenamefont {Tsirkin}, \citenamefont {Wierzbowska}, \citenamefont {Marzari}, \citenamefont {Vanderbilt}, \citenamefont {Souza}, \citenamefont {Mostofi},\ and\ \citenamefont {Yates}}]{Pizzi_2020}%
  \BibitemOpen
  \bibfield  {author} {\bibinfo {author} {\bibfnamefont {G.}~\bibnamefont {Pizzi}}, \bibinfo {author} {\bibfnamefont {V.}~\bibnamefont {Vitale}}, \bibinfo {author} {\bibfnamefont {R.}~\bibnamefont {Arita}}, \bibinfo {author} {\bibfnamefont {S.}~\bibnamefont {Blügel}}, \bibinfo {author} {\bibfnamefont {F.}~\bibnamefont {Freimuth}}, \bibinfo {author} {\bibfnamefont {G.}~\bibnamefont {Géranton}}, \bibinfo {author} {\bibfnamefont {M.}~\bibnamefont {Gibertini}}, \bibinfo {author} {\bibfnamefont {D.}~\bibnamefont {Gresch}}, \bibinfo {author} {\bibfnamefont {C.}~\bibnamefont {Johnson}}, \bibinfo {author} {\bibfnamefont {T.}~\bibnamefont {Koretsune}}, \bibinfo {author} {\bibfnamefont {J.}~\bibnamefont {Ibañez-Azpiroz}}, \bibinfo {author} {\bibfnamefont {H.}~\bibnamefont {Lee}}, \bibinfo {author} {\bibfnamefont {J.-M.}\ \bibnamefont {Lihm}}, \bibinfo {author} {\bibfnamefont {D.}~\bibnamefont {Marchand}}, \bibinfo {author} {\bibfnamefont {A.}~\bibnamefont {Marrazzo}}, \bibinfo {author} {\bibfnamefont
  {Y.}~\bibnamefont {Mokrousov}}, \bibinfo {author} {\bibfnamefont {J.~I.}\ \bibnamefont {Mustafa}}, \bibinfo {author} {\bibfnamefont {Y.}~\bibnamefont {Nohara}}, \bibinfo {author} {\bibfnamefont {Y.}~\bibnamefont {Nomura}}, \bibinfo {author} {\bibfnamefont {L.}~\bibnamefont {Paulatto}}, \bibinfo {author} {\bibfnamefont {S.}~\bibnamefont {Poncé}}, \bibinfo {author} {\bibfnamefont {T.}~\bibnamefont {Ponweiser}}, \bibinfo {author} {\bibfnamefont {J.}~\bibnamefont {Qiao}}, \bibinfo {author} {\bibfnamefont {F.}~\bibnamefont {Thöle}}, \bibinfo {author} {\bibfnamefont {S.~S.}\ \bibnamefont {Tsirkin}}, \bibinfo {author} {\bibfnamefont {M.}~\bibnamefont {Wierzbowska}}, \bibinfo {author} {\bibfnamefont {N.}~\bibnamefont {Marzari}}, \bibinfo {author} {\bibfnamefont {D.}~\bibnamefont {Vanderbilt}}, \bibinfo {author} {\bibfnamefont {I.}~\bibnamefont {Souza}}, \bibinfo {author} {\bibfnamefont {A.~A.}\ \bibnamefont {Mostofi}},\ and\ \bibinfo {author} {\bibfnamefont {J.~R.}\ \bibnamefont {Yates}},\ }\href
  {https://doi.org/10.1088/1361-648X/ab51ff} {\bibfield  {journal} {\bibinfo  {journal} {Journal of Physics: Condensed Matter}\ }\textbf {\bibinfo {volume} {32}},\ \bibinfo {pages} {165902} (\bibinfo {year} {2020})}\BibitemShut {NoStop}%
\bibitem [{\citenamefont {Jiang}\ \emph {et~al.}(2026)\citenamefont {Jiang}, \citenamefont {Petralanda}, \citenamefont {Pi}, \citenamefont {Skorupskii}, \citenamefont {Xu}, \citenamefont {C{\u{a}}lug{\u{a}}ru}, \citenamefont {Hu}, \citenamefont {Xie}, \citenamefont {Mustaf}, \citenamefont {H{\"o}hn} \emph {et~al.}}]{jiang20262d}%
  \BibitemOpen
  \bibfield  {author} {\bibinfo {author} {\bibfnamefont {Y.}~\bibnamefont {Jiang}}, \bibinfo {author} {\bibfnamefont {U.}~\bibnamefont {Petralanda}}, \bibinfo {author} {\bibfnamefont {H.}~\bibnamefont {Pi}}, \bibinfo {author} {\bibfnamefont {G.}~\bibnamefont {Skorupskii}}, \bibinfo {author} {\bibfnamefont {Q.}~\bibnamefont {Xu}}, \bibinfo {author} {\bibfnamefont {D.}~\bibnamefont {C{\u{a}}lug{\u{a}}ru}}, \bibinfo {author} {\bibfnamefont {H.}~\bibnamefont {Hu}}, \bibinfo {author} {\bibfnamefont {J.}~\bibnamefont {Xie}}, \bibinfo {author} {\bibfnamefont {R.~A.}\ \bibnamefont {Mustaf}}, \bibinfo {author} {\bibfnamefont {P.}~\bibnamefont {H{\"o}hn}}, \emph {et~al.},\ }\href@noop {} {\bibfield  {journal} {\bibinfo  {journal} {Science}\ }\textbf {\bibinfo {volume} {393}},\ \bibinfo {pages} {eadu1550} (\bibinfo {year} {2026})}\BibitemShut {NoStop}%
\bibitem [{\citenamefont {Ding}\ \emph {et~al.}(2025)\citenamefont {Ding}, \citenamefont {Verma},\ and\ \citenamefont {Queiroz}}]{ding2025cyclic}%
  \BibitemOpen
  \bibfield  {author} {\bibinfo {author} {\bibfnamefont {P.}~\bibnamefont {Ding}}, \bibinfo {author} {\bibfnamefont {N.}~\bibnamefont {Verma}},\ and\ \bibinfo {author} {\bibfnamefont {R.}~\bibnamefont {Queiroz}},\ }\href@noop {} {\bibfield  {journal} {\bibinfo  {journal} {arXiv preprint arXiv:2511.19634}\ } (\bibinfo {year} {2025})}\BibitemShut {NoStop}%
\bibitem [{\citenamefont {Ingham}\ and\ \citenamefont {Queiroz}(2026)}]{ingham2026real}%
  \BibitemOpen
  \bibfield  {author} {\bibinfo {author} {\bibfnamefont {J.}~\bibnamefont {Ingham}}\ and\ \bibinfo {author} {\bibfnamefont {R.}~\bibnamefont {Queiroz}},\ }\href@noop {} {\bibfield  {journal} {\bibinfo  {journal} {arXiv preprint arXiv:2607.21699}\ } (\bibinfo {year} {2026})}\BibitemShut {NoStop}%
\bibitem [{\citenamefont {Yu}\ \emph {et~al.}(2026)\citenamefont {Yu}, \citenamefont {Jiang}, \citenamefont {Xu}, \citenamefont {C{\u{a}}lug{\u{a}}ru}, \citenamefont {Hu}, \citenamefont {Guo}, \citenamefont {Sajan}, \citenamefont {Wang}, \citenamefont {Ugeda}, \citenamefont {De~Juan} \emph {et~al.}}]{yu2026quantum}%
  \BibitemOpen
  \bibfield  {author} {\bibinfo {author} {\bibfnamefont {J.}~\bibnamefont {Yu}}, \bibinfo {author} {\bibfnamefont {Y.}~\bibnamefont {Jiang}}, \bibinfo {author} {\bibfnamefont {Y.}~\bibnamefont {Xu}}, \bibinfo {author} {\bibfnamefont {D.}~\bibnamefont {C{\u{a}}lug{\u{a}}ru}}, \bibinfo {author} {\bibfnamefont {H.}~\bibnamefont {Hu}}, \bibinfo {author} {\bibfnamefont {H.}~\bibnamefont {Guo}}, \bibinfo {author} {\bibfnamefont {S.}~\bibnamefont {Sajan}}, \bibinfo {author} {\bibfnamefont {Y.}~\bibnamefont {Wang}}, \bibinfo {author} {\bibfnamefont {M.~M.}\ \bibnamefont {Ugeda}}, \bibinfo {author} {\bibfnamefont {F.}~\bibnamefont {De~Juan}}, \emph {et~al.},\ }\href@noop {} {\bibfield  {journal} {\bibinfo  {journal} {Physical Review B}\ }\textbf {\bibinfo {volume} {114}},\ \bibinfo {pages} {125104} (\bibinfo {year} {2026})}\BibitemShut {NoStop}%
\bibitem [{\citenamefont {C{\u{a}}lug{\u{a}}ru}\ \emph {et~al.}(2026)\citenamefont {C{\u{a}}lug{\u{a}}ru}, \citenamefont {Jiang}, \citenamefont {Guo}, \citenamefont {Sajan}, \citenamefont {Wang}, \citenamefont {Hu}, \citenamefont {Yu}, \citenamefont {Bernevig}, \citenamefont {de~Juan},\ and\ \citenamefont {Ugeda}}]{cualuguaru2026observation}%
  \BibitemOpen
  \bibfield  {author} {\bibinfo {author} {\bibfnamefont {D.}~\bibnamefont {C{\u{a}}lug{\u{a}}ru}}, \bibinfo {author} {\bibfnamefont {Y.}~\bibnamefont {Jiang}}, \bibinfo {author} {\bibfnamefont {H.}~\bibnamefont {Guo}}, \bibinfo {author} {\bibfnamefont {S.}~\bibnamefont {Sajan}}, \bibinfo {author} {\bibfnamefont {Y.}~\bibnamefont {Wang}}, \bibinfo {author} {\bibfnamefont {H.}~\bibnamefont {Hu}}, \bibinfo {author} {\bibfnamefont {J.}~\bibnamefont {Yu}}, \bibinfo {author} {\bibfnamefont {B.~A.}\ \bibnamefont {Bernevig}}, \bibinfo {author} {\bibfnamefont {F.}~\bibnamefont {de~Juan}},\ and\ \bibinfo {author} {\bibfnamefont {M.~M.}\ \bibnamefont {Ugeda}},\ }\href@noop {} {\bibfield  {journal} {\bibinfo  {journal} {Nature Physics}\ ,\ \bibinfo {pages} {1}} (\bibinfo {year} {2026})}\BibitemShut {NoStop}%
\bibitem [{\citenamefont {Bradlyn}\ \emph {et~al.}(2017)\citenamefont {Bradlyn}, \citenamefont {Elcoro}, \citenamefont {Cano}, \citenamefont {Vergniory}, \citenamefont {Wang}, \citenamefont {Felser}, \citenamefont {Aroyo},\ and\ \citenamefont {Bernevig}}]{bradlyn2017topological}%
  \BibitemOpen
  \bibfield  {author} {\bibinfo {author} {\bibfnamefont {B.}~\bibnamefont {Bradlyn}}, \bibinfo {author} {\bibfnamefont {L.}~\bibnamefont {Elcoro}}, \bibinfo {author} {\bibfnamefont {J.}~\bibnamefont {Cano}}, \bibinfo {author} {\bibfnamefont {M.~G.}\ \bibnamefont {Vergniory}}, \bibinfo {author} {\bibfnamefont {Z.}~\bibnamefont {Wang}}, \bibinfo {author} {\bibfnamefont {C.}~\bibnamefont {Felser}}, \bibinfo {author} {\bibfnamefont {M.~I.}\ \bibnamefont {Aroyo}},\ and\ \bibinfo {author} {\bibfnamefont {B.~A.}\ \bibnamefont {Bernevig}},\ }\href@noop {} {\bibfield  {journal} {\bibinfo  {journal} {Nature}\ }\textbf {\bibinfo {volume} {547}},\ \bibinfo {pages} {298} (\bibinfo {year} {2017})}\BibitemShut {NoStop}%
\bibitem [{\citenamefont {Cano}\ \emph {et~al.}(2018)\citenamefont {Cano}, \citenamefont {Bradlyn}, \citenamefont {Wang}, \citenamefont {Elcoro}, \citenamefont {Vergniory}, \citenamefont {Felser}, \citenamefont {Aroyo},\ and\ \citenamefont {Bernevig}}]{cano2018building}%
  \BibitemOpen
  \bibfield  {author} {\bibinfo {author} {\bibfnamefont {J.}~\bibnamefont {Cano}}, \bibinfo {author} {\bibfnamefont {B.}~\bibnamefont {Bradlyn}}, \bibinfo {author} {\bibfnamefont {Z.}~\bibnamefont {Wang}}, \bibinfo {author} {\bibfnamefont {L.}~\bibnamefont {Elcoro}}, \bibinfo {author} {\bibfnamefont {M.~G.}\ \bibnamefont {Vergniory}}, \bibinfo {author} {\bibfnamefont {C.}~\bibnamefont {Felser}}, \bibinfo {author} {\bibfnamefont {M.~I.}\ \bibnamefont {Aroyo}},\ and\ \bibinfo {author} {\bibfnamefont {B.~A.}\ \bibnamefont {Bernevig}},\ }\href@noop {} {\bibfield  {journal} {\bibinfo  {journal} {Phys. Rev. B}\ }\textbf {\bibinfo {volume} {97}},\ \bibinfo {pages} {035139} (\bibinfo {year} {2018})}\BibitemShut {NoStop}%
\bibitem [{\citenamefont {Xian}\ \emph {et~al.}(2021)\citenamefont {Xian}, \citenamefont {Claassen}, \citenamefont {Kiese}, \citenamefont {Scherer}, \citenamefont {Trebst}, \citenamefont {Kennes},\ and\ \citenamefont {Rubio}}]{xian2021realization}%
  \BibitemOpen
  \bibfield  {author} {\bibinfo {author} {\bibfnamefont {L.}~\bibnamefont {Xian}}, \bibinfo {author} {\bibfnamefont {M.}~\bibnamefont {Claassen}}, \bibinfo {author} {\bibfnamefont {D.}~\bibnamefont {Kiese}}, \bibinfo {author} {\bibfnamefont {M.~M.}\ \bibnamefont {Scherer}}, \bibinfo {author} {\bibfnamefont {S.}~\bibnamefont {Trebst}}, \bibinfo {author} {\bibfnamefont {D.~M.}\ \bibnamefont {Kennes}},\ and\ \bibinfo {author} {\bibfnamefont {A.}~\bibnamefont {Rubio}},\ }\href@noop {} {\bibfield  {journal} {\bibinfo  {journal} {Nature communications}\ }\textbf {\bibinfo {volume} {12}},\ \bibinfo {pages} {5644} (\bibinfo {year} {2021})}\BibitemShut {NoStop}%
\bibitem [{\citenamefont {Pan}\ \emph {et~al.}(2020)\citenamefont {Pan}, \citenamefont {Wu},\ and\ \citenamefont {Das~Sarma}}]{pan2020band}%
  \BibitemOpen
  \bibfield  {author} {\bibinfo {author} {\bibfnamefont {H.}~\bibnamefont {Pan}}, \bibinfo {author} {\bibfnamefont {F.}~\bibnamefont {Wu}},\ and\ \bibinfo {author} {\bibfnamefont {S.}~\bibnamefont {Das~Sarma}},\ }\href {https://doi.org/10.1103/PhysRevResearch.2.033087} {\bibfield  {journal} {\bibinfo  {journal} {Physical Review Research}\ }\textbf {\bibinfo {volume} {2}},\ \bibinfo {pages} {033087} (\bibinfo {year} {2020})}\BibitemShut {NoStop}%
\bibitem [{\citenamefont {Brzezi\'nska}\ \emph {et~al.}(2024)\citenamefont {Brzezi\'nska}, \citenamefont {Grytsiuk}, \citenamefont {R\"osner}, \citenamefont {Gibertini},\ and\ \citenamefont {Rademaker}}]{brzezinska2024pressure}%
  \BibitemOpen
  \bibfield  {author} {\bibinfo {author} {\bibfnamefont {M.}~\bibnamefont {Brzezi\'nska}}, \bibinfo {author} {\bibfnamefont {S.}~\bibnamefont {Grytsiuk}}, \bibinfo {author} {\bibfnamefont {M.}~\bibnamefont {R\"osner}}, \bibinfo {author} {\bibfnamefont {M.}~\bibnamefont {Gibertini}},\ and\ \bibinfo {author} {\bibfnamefont {L.}~\bibnamefont {Rademaker}},\ }\href {https://doi.org/10.1088/2053-1583/ad7c5f} {\bibfield  {journal} {\bibinfo  {journal} {2D Materials}\ }\textbf {\bibinfo {volume} {12}},\ \bibinfo {pages} {015003} (\bibinfo {year} {2024})}\BibitemShut {NoStop}%
\bibitem [{\citenamefont {Hu}\ \emph {et~al.}(2022)\citenamefont {Hu}, \citenamefont {Zhang}, \citenamefont {Xie},\ and\ \citenamefont {Law}}]{hu2022nonlinear}%
  \BibitemOpen
  \bibfield  {author} {\bibinfo {author} {\bibfnamefont {J.-X.}\ \bibnamefont {Hu}}, \bibinfo {author} {\bibfnamefont {C.-P.}\ \bibnamefont {Zhang}}, \bibinfo {author} {\bibfnamefont {Y.-M.}\ \bibnamefont {Xie}},\ and\ \bibinfo {author} {\bibfnamefont {K.}~\bibnamefont {Law}},\ }\href@noop {} {\bibfield  {journal} {\bibinfo  {journal} {Communications Physics}\ }\textbf {\bibinfo {volume} {5}},\ \bibinfo {pages} {255} (\bibinfo {year} {2022})}\BibitemShut {NoStop}%
\bibitem [{\citenamefont {Foutty}\ \emph {et~al.}(2024)\citenamefont {Foutty}, \citenamefont {Kometter}, \citenamefont {Devakul}, \citenamefont {Reddy}, \citenamefont {Watanabe}, \citenamefont {Taniguchi}, \citenamefont {Fu},\ and\ \citenamefont {Feldman}}]{foutty2024mapping}%
  \BibitemOpen
  \bibfield  {author} {\bibinfo {author} {\bibfnamefont {B.~A.}\ \bibnamefont {Foutty}}, \bibinfo {author} {\bibfnamefont {C.~R.}\ \bibnamefont {Kometter}}, \bibinfo {author} {\bibfnamefont {T.}~\bibnamefont {Devakul}}, \bibinfo {author} {\bibfnamefont {A.~P.}\ \bibnamefont {Reddy}}, \bibinfo {author} {\bibfnamefont {K.}~\bibnamefont {Watanabe}}, \bibinfo {author} {\bibfnamefont {T.}~\bibnamefont {Taniguchi}}, \bibinfo {author} {\bibfnamefont {L.}~\bibnamefont {Fu}},\ and\ \bibinfo {author} {\bibfnamefont {B.~E.}\ \bibnamefont {Feldman}},\ }\href {https://doi.org/10.1126/science.adi4728} {\bibfield  {journal} {\bibinfo  {journal} {Science}\ }\textbf {\bibinfo {volume} {384}},\ \bibinfo {pages} {343} (\bibinfo {year} {2024})}\BibitemShut {NoStop}%
\bibitem [{\citenamefont {Kang}\ \emph {et~al.}(2024)\citenamefont {Kang}, \citenamefont {Qiu}, \citenamefont {Watanabe}, \citenamefont {Taniguchi}, \citenamefont {Shan},\ and\ \citenamefont {Mak}}]{kang2024double}%
  \BibitemOpen
  \bibfield  {author} {\bibinfo {author} {\bibfnamefont {K.}~\bibnamefont {Kang}}, \bibinfo {author} {\bibfnamefont {Y.}~\bibnamefont {Qiu}}, \bibinfo {author} {\bibfnamefont {K.}~\bibnamefont {Watanabe}}, \bibinfo {author} {\bibfnamefont {T.}~\bibnamefont {Taniguchi}}, \bibinfo {author} {\bibfnamefont {J.}~\bibnamefont {Shan}},\ and\ \bibinfo {author} {\bibfnamefont {K.~F.}\ \bibnamefont {Mak}},\ }\href {https://doi.org/10.1021/acs.nanolett.4c05308} {\bibfield  {journal} {\bibinfo  {journal} {Nano Letters}\ }\textbf {\bibinfo {volume} {24}},\ \bibinfo {pages} {14901} (\bibinfo {year} {2024})}\BibitemShut {NoStop}%
\bibitem [{\citenamefont {Gao}\ \emph {et~al.}(2026)\citenamefont {Gao}, \citenamefont {Ghafariasl}, \citenamefont {Mehrabad}, \citenamefont {Huang}, \citenamefont {Zhang}, \citenamefont {Session}, \citenamefont {Upadhyay}, \citenamefont {Ma}, \citenamefont {Alshalan}, \citenamefont {Suarez}, \citenamefont {Sarkar}, \citenamefont {Park}, \citenamefont {Jang}, \citenamefont {Watanabe}, \citenamefont {Taniguchi}, \citenamefont {Xie}, \citenamefont {Zhou},\ and\ \citenamefont {Hafezi}}]{gao2026probing}%
  \BibitemOpen
  \bibfield  {author} {\bibinfo {author} {\bibfnamefont {B.}~\bibnamefont {Gao}}, \bibinfo {author} {\bibfnamefont {M.}~\bibnamefont {Ghafariasl}}, \bibinfo {author} {\bibfnamefont {M.~J.}\ \bibnamefont {Mehrabad}}, \bibinfo {author} {\bibfnamefont {T.-S.}\ \bibnamefont {Huang}}, \bibinfo {author} {\bibfnamefont {L.}~\bibnamefont {Zhang}}, \bibinfo {author} {\bibfnamefont {D.}~\bibnamefont {Session}}, \bibinfo {author} {\bibfnamefont {P.}~\bibnamefont {Upadhyay}}, \bibinfo {author} {\bibfnamefont {R.}~\bibnamefont {Ma}}, \bibinfo {author} {\bibfnamefont {G.}~\bibnamefont {Alshalan}}, \bibinfo {author} {\bibfnamefont {D.}~\bibnamefont {Suarez}}, \bibinfo {author} {\bibfnamefont {S.}~\bibnamefont {Sarkar}}, \bibinfo {author} {\bibfnamefont {S.}~\bibnamefont {Park}}, \bibinfo {author} {\bibfnamefont {H.}~\bibnamefont {Jang}}, \bibinfo {author} {\bibfnamefont {K.}~\bibnamefont {Watanabe}}, \bibinfo {author} {\bibfnamefont {T.}~\bibnamefont {Taniguchi}}, \bibinfo {author} {\bibfnamefont {M.}~\bibnamefont {Xie}},
  \bibinfo {author} {\bibfnamefont {Y.}~\bibnamefont {Zhou}},\ and\ \bibinfo {author} {\bibfnamefont {M.}~\bibnamefont {Hafezi}},\ }\href {https://doi.org/10.1103/259w-bpxx} {\bibfield  {journal} {\bibinfo  {journal} {Physical Review X}\ }\textbf {\bibinfo {volume} {16}},\ \bibinfo {pages} {021037} (\bibinfo {year} {2026})}\BibitemShut {NoStop}%
\bibitem [{\citenamefont {Wang}\ \emph {et~al.}(2020)\citenamefont {Wang}, \citenamefont {Shih}, \citenamefont {Ghiotto}, \citenamefont {Xian}, \citenamefont {Rhodes}, \citenamefont {Tan}, \citenamefont {Claassen}, \citenamefont {Kennes}, \citenamefont {Bai}, \citenamefont {Kim} \emph {et~al.}}]{wang2020correlated}%
  \BibitemOpen
  \bibfield  {author} {\bibinfo {author} {\bibfnamefont {L.}~\bibnamefont {Wang}}, \bibinfo {author} {\bibfnamefont {E.-M.}\ \bibnamefont {Shih}}, \bibinfo {author} {\bibfnamefont {A.}~\bibnamefont {Ghiotto}}, \bibinfo {author} {\bibfnamefont {L.}~\bibnamefont {Xian}}, \bibinfo {author} {\bibfnamefont {D.~A.}\ \bibnamefont {Rhodes}}, \bibinfo {author} {\bibfnamefont {C.}~\bibnamefont {Tan}}, \bibinfo {author} {\bibfnamefont {M.}~\bibnamefont {Claassen}}, \bibinfo {author} {\bibfnamefont {D.~M.}\ \bibnamefont {Kennes}}, \bibinfo {author} {\bibfnamefont {Y.}~\bibnamefont {Bai}}, \bibinfo {author} {\bibfnamefont {B.}~\bibnamefont {Kim}}, \emph {et~al.},\ }\href@noop {} {\bibfield  {journal} {\bibinfo  {journal} {Nature Materials}\ }\textbf {\bibinfo {volume} {19}},\ \bibinfo {pages} {861} (\bibinfo {year} {2020})}\BibitemShut {NoStop}%
\bibitem [{\citenamefont {Xia}\ \emph {et~al.}(2026{\natexlab{a}})\citenamefont {Xia}, \citenamefont {Han}, \citenamefont {Zhu}, \citenamefont {Zhang}, \citenamefont {Kn{\"u}ppel}, \citenamefont {Watanabe}, \citenamefont {Taniguchi}, \citenamefont {Mak},\ and\ \citenamefont {Shan}}]{xia2026bandwidth}%
  \BibitemOpen
  \bibfield  {author} {\bibinfo {author} {\bibfnamefont {Y.}~\bibnamefont {Xia}}, \bibinfo {author} {\bibfnamefont {Z.}~\bibnamefont {Han}}, \bibinfo {author} {\bibfnamefont {J.}~\bibnamefont {Zhu}}, \bibinfo {author} {\bibfnamefont {v.}~\bibnamefont {Zhang}}, \bibinfo {author} {\bibfnamefont {P.}~\bibnamefont {Kn{\"u}ppel}}, \bibinfo {author} {\bibfnamefont {K.}~\bibnamefont {Watanabe}}, \bibinfo {author} {\bibfnamefont {T.}~\bibnamefont {Taniguchi}}, \bibinfo {author} {\bibfnamefont {K.~F.}\ \bibnamefont {Mak}},\ and\ \bibinfo {author} {\bibfnamefont {J.}~\bibnamefont {Shan}},\ }\href@noop {} {\bibfield  {journal} {\bibinfo  {journal} {Nature}\ }\textbf {\bibinfo {volume} {650}},\ \bibinfo {pages} {585} (\bibinfo {year} {2026}{\natexlab{a}})}\BibitemShut {NoStop}%
\bibitem [{\citenamefont {Xia}\ \emph {et~al.}(2026{\natexlab{b}})\citenamefont {Xia}, \citenamefont {Zhang}, \citenamefont {Han}, \citenamefont {Li}, \citenamefont {Bauer}, \citenamefont {Watanabe}, \citenamefont {Taniguchi}, \citenamefont {Zhao}, \citenamefont {Shan},\ and\ \citenamefont {Mak}}]{xia2026extended}%
  \BibitemOpen
  \bibfield  {author} {\bibinfo {author} {\bibfnamefont {Y.}~\bibnamefont {Xia}}, \bibinfo {author} {\bibfnamefont {Y.}~\bibnamefont {Zhang}}, \bibinfo {author} {\bibfnamefont {Z.}~\bibnamefont {Han}}, \bibinfo {author} {\bibfnamefont {H.}~\bibnamefont {Li}}, \bibinfo {author} {\bibfnamefont {B.}~\bibnamefont {Bauer}}, \bibinfo {author} {\bibfnamefont {K.}~\bibnamefont {Watanabe}}, \bibinfo {author} {\bibfnamefont {T.}~\bibnamefont {Taniguchi}}, \bibinfo {author} {\bibfnamefont {W.}~\bibnamefont {Zhao}}, \bibinfo {author} {\bibfnamefont {J.}~\bibnamefont {Shan}},\ and\ \bibinfo {author} {\bibfnamefont {K.~F.}\ \bibnamefont {Mak}},\ }\href@noop {} {\bibfield  {journal} {\bibinfo  {journal} {arXiv preprint arXiv:2610.22307}\ } (\bibinfo {year} {2026}{\natexlab{b}})}\BibitemShut {NoStop}%
\bibitem [{\citenamefont {Li}\ \emph {et~al.}(2025)\citenamefont {Li}, \citenamefont {Bernevig},\ and\ \citenamefont {Regnault}}]{li2025multiband}%
  \BibitemOpen
  \bibfield  {author} {\bibinfo {author} {\bibfnamefont {H.}~\bibnamefont {Li}}, \bibinfo {author} {\bibfnamefont {B.~A.}\ \bibnamefont {Bernevig}},\ and\ \bibinfo {author} {\bibfnamefont {N.}~\bibnamefont {Regnault}},\ }\href@noop {} {\bibfield  {journal} {\bibinfo  {journal} {Physical Review B}\ }\textbf {\bibinfo {volume} {112}},\ \bibinfo {pages} {075130} (\bibinfo {year} {2025})}\BibitemShut {NoStop}%
\bibitem [{\citenamefont {Wang}\ \emph {et~al.}(2025)\citenamefont {Wang}, \citenamefont {Clark}, \citenamefont {Das}, \citenamefont {Zhu}, \citenamefont {Massatt}, \citenamefont {Gavini}, \citenamefont {Luskin},\ and\ \citenamefont {Ortner}}]{wang2025atomic}%
  \BibitemOpen
  \bibfield  {author} {\bibinfo {author} {\bibfnamefont {Y.}~\bibnamefont {Wang}}, \bibinfo {author} {\bibfnamefont {D.}~\bibnamefont {Clark}}, \bibinfo {author} {\bibfnamefont {S.}~\bibnamefont {Das}}, \bibinfo {author} {\bibfnamefont {Z.}~\bibnamefont {Zhu}}, \bibinfo {author} {\bibfnamefont {D.}~\bibnamefont {Massatt}}, \bibinfo {author} {\bibfnamefont {V.}~\bibnamefont {Gavini}}, \bibinfo {author} {\bibfnamefont {M.}~\bibnamefont {Luskin}},\ and\ \bibinfo {author} {\bibfnamefont {C.}~\bibnamefont {Ortner}},\ }\href@noop {} {\bibfield  {journal} {\bibinfo  {journal} {Machine Learning: Science and Technology}\ }\textbf {\bibinfo {volume} {6}},\ \bibinfo {pages} {045040} (\bibinfo {year} {2025})}\BibitemShut {NoStop}%
\bibitem [{\citenamefont {Giannozzi}\ \emph {et~al.}(2009)\citenamefont {Giannozzi}, \citenamefont {Baroni}, \citenamefont {Bonini}, \citenamefont {Calandra}, \citenamefont {Car}, \citenamefont {Cavazzoni}, \citenamefont {Ceresoli}, \citenamefont {Chiarotti}, \citenamefont {Cococcioni}, \citenamefont {Dabo} \emph {et~al.}}]{giannozzi2009quantum}%
  \BibitemOpen
  \bibfield  {author} {\bibinfo {author} {\bibfnamefont {P.}~\bibnamefont {Giannozzi}}, \bibinfo {author} {\bibfnamefont {S.}~\bibnamefont {Baroni}}, \bibinfo {author} {\bibfnamefont {N.}~\bibnamefont {Bonini}}, \bibinfo {author} {\bibfnamefont {M.}~\bibnamefont {Calandra}}, \bibinfo {author} {\bibfnamefont {R.}~\bibnamefont {Car}}, \bibinfo {author} {\bibfnamefont {C.}~\bibnamefont {Cavazzoni}}, \bibinfo {author} {\bibfnamefont {D.}~\bibnamefont {Ceresoli}}, \bibinfo {author} {\bibfnamefont {G.~L.}\ \bibnamefont {Chiarotti}}, \bibinfo {author} {\bibfnamefont {M.}~\bibnamefont {Cococcioni}}, \bibinfo {author} {\bibfnamefont {I.}~\bibnamefont {Dabo}}, \emph {et~al.},\ }\href@noop {} {\bibfield  {journal} {\bibinfo  {journal} {Journal of physics: Condensed matter}\ }\textbf {\bibinfo {volume} {21}},\ \bibinfo {pages} {395502} (\bibinfo {year} {2009})}\BibitemShut {NoStop}%
\bibitem [{\citenamefont {Hamann}(2013)}]{hamann2013optimized}%
  \BibitemOpen
  \bibfield  {author} {\bibinfo {author} {\bibfnamefont {D.~R.}\ \bibnamefont {Hamann}},\ }\href@noop {} {\bibfield  {journal} {\bibinfo  {journal} {Physical Review B—Condensed Matter and Materials Physics}\ }\textbf {\bibinfo {volume} {88}},\ \bibinfo {pages} {085117} (\bibinfo {year} {2013})}\BibitemShut {NoStop}%
\bibitem [{\citenamefont {Van~Setten}\ \emph {et~al.}(2018)\citenamefont {Van~Setten}, \citenamefont {Giantomassi}, \citenamefont {Bousquet}, \citenamefont {Verstraete}, \citenamefont {Hamann}, \citenamefont {Gonze},\ and\ \citenamefont {Rignanese}}]{van2018pseudodojo}%
  \BibitemOpen
  \bibfield  {author} {\bibinfo {author} {\bibfnamefont {M.~J.}\ \bibnamefont {Van~Setten}}, \bibinfo {author} {\bibfnamefont {M.}~\bibnamefont {Giantomassi}}, \bibinfo {author} {\bibfnamefont {E.}~\bibnamefont {Bousquet}}, \bibinfo {author} {\bibfnamefont {M.~J.}\ \bibnamefont {Verstraete}}, \bibinfo {author} {\bibfnamefont {D.~R.}\ \bibnamefont {Hamann}}, \bibinfo {author} {\bibfnamefont {X.}~\bibnamefont {Gonze}},\ and\ \bibinfo {author} {\bibfnamefont {G.-M.}\ \bibnamefont {Rignanese}},\ }\href@noop {} {\bibfield  {journal} {\bibinfo  {journal} {Computer Physics Communications}\ }\textbf {\bibinfo {volume} {226}},\ \bibinfo {pages} {39} (\bibinfo {year} {2018})}\BibitemShut {NoStop}%
\bibitem [{\citenamefont {Mostofi}\ \emph {et~al.}(2014)\citenamefont {Mostofi}, \citenamefont {Yates}, \citenamefont {Pizzi}, \citenamefont {Lee}, \citenamefont {Souza}, \citenamefont {Vanderbilt},\ and\ \citenamefont {Marzari}}]{mostofi2014updated}%
  \BibitemOpen
  \bibfield  {author} {\bibinfo {author} {\bibfnamefont {A.~A.}\ \bibnamefont {Mostofi}}, \bibinfo {author} {\bibfnamefont {J.~R.}\ \bibnamefont {Yates}}, \bibinfo {author} {\bibfnamefont {G.}~\bibnamefont {Pizzi}}, \bibinfo {author} {\bibfnamefont {Y.-S.}\ \bibnamefont {Lee}}, \bibinfo {author} {\bibfnamefont {I.}~\bibnamefont {Souza}}, \bibinfo {author} {\bibfnamefont {D.}~\bibnamefont {Vanderbilt}},\ and\ \bibinfo {author} {\bibfnamefont {N.}~\bibnamefont {Marzari}},\ }\href@noop {} {\bibfield  {journal} {\bibinfo  {journal} {Computer Physics Communications}\ }\textbf {\bibinfo {volume} {185}},\ \bibinfo {pages} {2309} (\bibinfo {year} {2014})}\BibitemShut {NoStop}%
\bibitem [{\citenamefont {Kawamura}\ \emph {et~al.}(2014)\citenamefont {Kawamura}, \citenamefont {Gohda},\ and\ \citenamefont {Tsuneyuki}}]{kawamura2014improved}%
  \BibitemOpen
  \bibfield  {author} {\bibinfo {author} {\bibfnamefont {M.}~\bibnamefont {Kawamura}}, \bibinfo {author} {\bibfnamefont {Y.}~\bibnamefont {Gohda}},\ and\ \bibinfo {author} {\bibfnamefont {S.}~\bibnamefont {Tsuneyuki}},\ }\href {https://doi.org/10.1103/PhysRevB.89.094515} {\bibfield  {journal} {\bibinfo  {journal} {Physical Review B}\ }\textbf {\bibinfo {volume} {89}},\ \bibinfo {pages} {094515} (\bibinfo {year} {2014})}\BibitemShut {NoStop}%
\bibitem [{\citenamefont {Reed}\ and\ \citenamefont {Simon}(1975)}]{reed1975ii}%
  \BibitemOpen
  \bibfield  {author} {\bibinfo {author} {\bibfnamefont {M.}~\bibnamefont {Reed}}\ and\ \bibinfo {author} {\bibfnamefont {B.}~\bibnamefont {Simon}},\ }\href@noop {} {\emph {\bibinfo {title} {II: Fourier analysis, self-adjointness}}},\ Vol.~\bibinfo {volume} {2}\ (\bibinfo  {publisher} {Elsevier},\ \bibinfo {year} {1975})\BibitemShut {NoStop}%
\bibitem [{\citenamefont {Katznelson}(1968)}]{katznelson1968introduction}%
  \BibitemOpen
  \bibfield  {author} {\bibinfo {author} {\bibfnamefont {Y.}~\bibnamefont {Katznelson}},\ }\href@noop {} {\emph {\bibinfo {title} {An introduction to harmonic analysis}}}\ (\bibinfo  {publisher} {Wiley New York, London, Sydney Toronto},\ \bibinfo {year} {1968})\BibitemShut {NoStop}%
\end{thebibliography}%

\appendix
\setcounter{figure}{0}
\renewcommand{\thefigure}{S\arabic{figure}}

\section{Computational details} \label{sec:code}

First-principles density functional theory (DFT) calculations were performed using the Quantum ESPRESSO package~\cite{giannozzi2009quantum} (version 7.3.1). Exchange-correlation effects were treated within the generalized gradient approximation using the Perdew-Burke-Ernzerhof exchange-correlation functional. We used fully relativistic, optimized norm-conserving Vanderbilt pseudopotentials~\cite{hamann2013optimized} obtained from Pseudo-Dojo~\cite{van2018pseudodojo}, with plane-wave kinetic-energy and charge-density cutoffs of $80$ Ry and $800$ Ry, respectively. Van der Waals interactions between layers were included through the Grimme-D3 dispersion correction. We first relaxed the monolayer structure in both lattice constants and atomic coordinates in all directions. 

\begin{figure}
    \centering  \includegraphics[width=\linewidth]{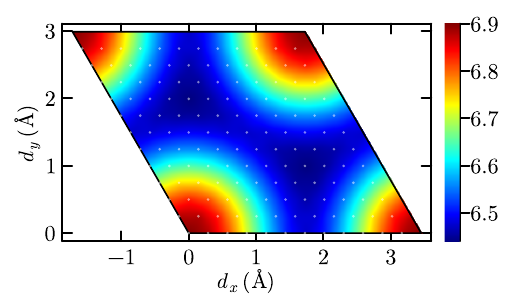}
    \caption{Interlayer Nb-Nb distance (\AA) in bilayer NbSe$_2$ as a function of in-plane translation $d$ of the top layer relative to the bottom layer.}
    \label{fig:interlayer_distance}
\end{figure}

For the commensurate bilayers considered, we fixed the in-plane lattice constant to be the same as that obtained in the monolayer and relaxed independently for each $d$-shift only along the out-of-plane direction using a $\Gamma$-centered ${k}$-grid of $20\times 20 \times 1$ with a force convergence threshold of $10^{-3} \, \mathrm{Ry}/\text{Bohr}$. Self-consistent calculations were carried out on a $\Gamma$-centered ${k}$-grid of $25\times 25 \times 1$ with a convergence threshold of $10^{-8}$ Ry. The non-self-consistent step in the DFT calculations was performed on a ${k}$-grid of $20\times 20 \times 1$. As a post-processing step, we constructed four single-shot Wannier functions that preserve the crystal symmetry using the Wannier90 code package~\cite{mostofi2014updated,Pizzi_2020}. 

For the twisted bilayer at $\theta = 9.43^\circ$, we performed a DFT calculation of the full $(m,n)=(3,4)$ commensurate moir\'e cell ($222$ atoms, $a_M = 20.99$~\AA) with the same functional, pseudopotentials, cutoffs and dispersion correction, relaxing all atomic positions at the $\Gamma$ point with the monolayer lattice constant fixed (force threshold $10^{-3}\,\mathrm{Ry}/\text{Bohr}$). The self-consistent calculation including spin-orbit coupling was carried out on a $6\times 6\times 1$ moir\'e ${k}$-grid with $0.005$ Ry Gaussian smearing, followed by a non-self-consistent calculation on a $12\times 12\times 1$ grid with the optimized tetrahedron method~\cite{kawamura2014improved}. The resulting density of states is convolved with a $15$ meV Gaussian and normalized per primitive bilayer cell.

\section{Asymptotics of the tunneling function} \label{app:decayharmonics}

This appendix reviews and adapts arguments used in Ref.~\cite{watson2023bistritzer} for the decay of the tunneling amplitudes in momentum space within the two-center approximation.
Our goal is to make these arguments more accessible and to show that, applied to the registry function $\bar t_{\beta\beta'}$ of Sec.~\ref{sec:wannier}, they lead to different scaling laws for the decay in the relative momentum $q$ and in the registry harmonics $\bar G$. Through the identification of Eq.~\eqref{eq:registryhyp_FT}, these laws carry over to the coefficients $t_{\beta\beta'}(q;g)$ of the environment-resolved continuum model, and fix the hierarchy of terms in Eq.~\eqref{eq:momentum_moire_tunnelings_infull}.
We first illustrate the mechanism on a minimal example, then review the mathematical argument ensuring its robustness, and end with the consequences for this hierarchy for the terms in Eq.~\eqref{eq:momentum_moire_tunnelings_infull}.

\subsection{Illustrative example for the relative coordinate}

Consider first the dependence on the relative coordinate $\bar\delta$. Since it couples exponentially localized Wannier states, the tunneling decays exponentially for large separation, $e^{-\kappa |\bar\delta|}$ with $\kappa^{-1}$ the localization length, and saturates smoothly to a constant $t_{0}$ at short distances. Both properties are captured by the simple model
\begin{equation}\label{eq:regkernel}
\bar t(\bar\delta) = t_{0}\, e^{ -\kappa \left( \sqrt{\bar\delta^{2}+\varepsilon^{2}} - \varepsilon \right)},
\end{equation}
whose two-dimensional Fourier transform reads
\begin{equation}\label{eq:regFT}
\bar t(q) = \frac{2\pi \kappa\, t_{0}}{\Abz}\, \frac{(1+\varepsilon u)\, e^{\varepsilon(\kappa-u)}}{u^{3}}, \quad u=\sqrt{\kappa^{2}+q^{2}} .
\end{equation}
Here, $\varepsilon$ is a short-distance (UV) regularizer representing the shortest physically relevant length scale: it is of order the monolayer lattice constant $a$ for intralayer processes, and of order the mean interlayer distance $h_z$ for interlayer ones.

Eq.~\eqref{eq:regFT} exhibits three regimes: (1) a smooth plateau for $q\lesssim\kappa$; (2) an approximately algebraic decay $\propto q^{-3}$ for $\kappa \lesssim q \lesssim \varepsilon^{-1}$, a window that exists only when $\varepsilon\kappa\lesssim1$; (3) an exponential tail $\sim e^{-\varepsilon q}/q^{2}$ for $\varepsilon q\gtrsim 1$. The important feature is not the specific form of Eq.~\eqref{eq:regkernel}, but the coexistence of the two microscopic scales $\kappa^{-1}$ and $\varepsilon$, which produces a broad algebraic regime before an exponential cutoff at large momenta.
Since the identification Eq.~\eqref{eq:registryhyp_FT} only rotates the momentum, $|L_{\ell\ell'}^{-1}q| = |q|$, these scaling laws in $q$ also apply to the continuum coefficients $t_{\beta\beta'}(q;g)$.

\subsection{Extension to center of mass coordinate}

The situation is qualitatively different for the registry coordinate. While the relative coordinate probes the spatial extent of the Wannier orbitals, the registry dependence reflects how the local environment varies with stacking, through changes in interlayer tunneling pathways, electrostatic potentials, and related effects. Yet, the large-momentum exponential decay can still be derived, largely independent of the exact functional form that these contributions take.
As pointed out in Ref.~\cite{watson2023bistritzer}, the relevant mathematical result is that Fourier coefficients decay exponentially whenever the underlying function admits an analytic continuation into a strip of the complex plane, with the decay rate ($\varepsilon$ above) determined by the distance to the nearest singularity~\cite[Th.~IX-14]{reed1975ii}. For periodic functions, the analogous statement is given in Ref.~\cite[Ex.~4, Sec.~I.4]{katznelson1968introduction}.

For the relative coordinate, the only physical singularity for tunneling between orbitals occurs when the physical distance $s=\sqrt{\bar\delta^2+\varepsilon^{2}}$ vanishes (\textit{i.e.} when the orbitals of the lattice collapse). Due to the regularizer, this can only occur after analytic continuation into the complex plane $\bar\delta \to i\eta$ for $\eta \simeq \pm \varepsilon$, and by the previous theorem, yields the exponential tail of Eq.~\eqref{eq:regFT} for $\varepsilon q\gtrsim1$.

For the registry coordinate, the same argument remains true irrespective of whether the underlying microscopic mechanism is tunneling-mediated or electrostatic in nature.
Indeed, every registry-dependent contribution describes a stacking effect and depends on the true three-dimensional distance between constituents in the different layers, $s=\sqrt{|\bar\delta + d|^{2}+h_z^{2}}$ with $h_z$ the interlayer distance. Following the same reasoning, the registry harmonics decay as $\bar t_{\beta\beta'}(q;\bar{G}) \sim e^{-h_z|\bar{G}|}$ for $|\bar{G}| a \gtrsim 1$.
Through the identification $t_{\beta\beta'}(q;g)=\bar t_{\beta\beta'}(L_{\ell\ell'}^{-1}q;\bar G_g)$ of Eq.~\eqref{eq:registryhyp_FT}, and noting that $|\bar{G}_g| = |\mathcal L^{-T} g| \sim |g| / \theta$ for small angles, this translates into a decay of the continuum harmonics on the moiré scale: $t_{\beta\beta'}(q; g) \sim e^{-\varepsilon_{m}|g|}$ when $|g| a_m \gtrsim1$, with $\varepsilon_{m} \sim h_z / \theta$ satisfying $a \ll \varepsilon_{m} \lesssim a_m$.

\subsection{Scaling laws: algebraic in $q$, exponential in $g$}

The hierarchy of terms in Eq.~\eqref{eq:momentum_moire_tunnelings_infull} comes from these the scaling laws. 
Writing $G_\ell = L_\ell^{-T} \bar G_{1}$ and $G_{\ell'} = L_{\ell'}^{-T} \bar G_{2}$, the delta-function constraint of Eq.~\eqref{eq:momentum_moire_tunnelings_infull} with $k$ and $k'$ separated by a few moiré reciprocal vectors requires the registry harmonic to absorb an atomic-scale momentum, $g \simeq \bar G_{2}-\bar G_{1}$, whenever the two are not resonant, $\bar G_{1} \neq\bar G_{2}$.
The corresponding coefficient $t_{\beta\beta'}(q; g)$ is suppressed by the factor $e^{-\varepsilon_m |g|}$ which, since the separation between monolayer and moiré length scales ensures $\varepsilon_m |\bar G_{2} - \bar G_{1}| \sim \theta^{-1} \gg 1$, is exponentially small beyond all orders in the twist angle. As a result, only the resonant channels $\bar G_1=\bar G_2\equiv\bar G$ survive; the same estimate bounds the commensuration (lock-in) effects neglected when assuming moiré periodicity in Sec.~\ref{ssec:ERTCA}.

\begin{figure}
    \centering  \includegraphics[width=\linewidth]{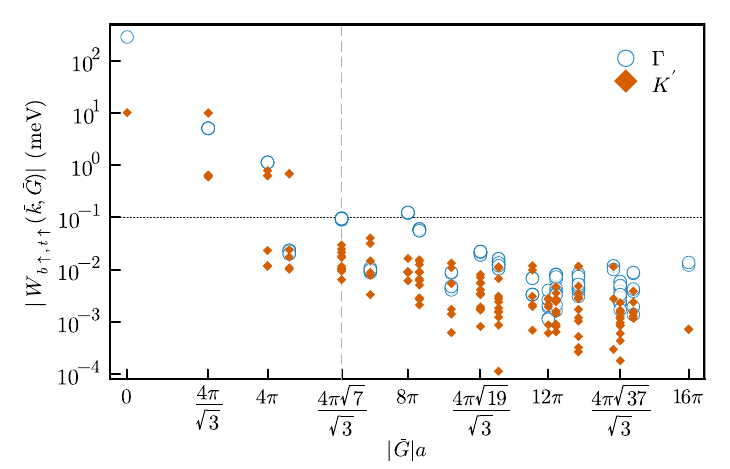}
    \caption{Decay of spin-conserving interlayer coefficients $W_{b\uparrow,t\uparrow}(\bar{k},\bar{G})$ as a function of $|\bar{G}|\,a$ in bilayer NbSe$_2$. The vertical grey dashed line indicates the fourth shell of harmonics, and the black horizontal dotted line indicates 0.1 meV.}
    \label{fig:harmonic_decay}
\end{figure}

The remaining resonant contributions differ only through their relative momentum shifted by $\bar G$. They therefore organize into shells whose amplitudes decrease according to the algebraic hierarchy inherited from Eq.~\eqref{eq:regFT}, with the eventual exponential cutoff setting in only at sufficiently large $|\bar G|$.
This distinction underlies the structure of the final formula of Sec.~\ref{sec:wannier}: the non-resonant channels are discarded because they are exponentially suppressed in the registry harmonics, while the resonant shells -- whose amplitudes decay only algebraically before crossing over to an exponential regime -- are resummed exactly into the Wannier harmonics $W_{\beta\beta'}$ of Eq.~\eqref{eq:DFTtoContinuum}, the sum over $\bar H$ being carried out automatically by the $d$-scan. This slow initial decay explains why second-harmonic terms, automatically included in our construction, are frequently required for a quantitatively accurate description of moiré~\cite{jia2024moire,mao2024transfer,zhang2024universal}. As an illustration, Fig.~\ref{fig:harmonic_decay} presents the interlayer Wannier harmonics of bilayer NbSe$_2$ as a function of $|\bar G|$  at $K'$ and $\Gamma$. Their amplitudes decay exponentially up to the fourth harmonic shell (vertical grey dashed line) and remain below 0.1 meV thereafter, bordering DFT numerical precision. We therefore retain only the first four harmonic shells throughout this work.

\section{Fourier transform and gauge conventions} \label{app_FTconvention}

This appendix summarizes the basis and Fourier transform conventions used to relate commensurate bilayer calculations to the continuum coefficients $\bar{t}_{\beta\beta'}(q;\bar G)$ found in Eq.~(\ref{eq:registryhyp_FT}).

\subsection{Absolute-position and Wannier bases}
For a commensurate bilayer with relative translation $d$, the absolute-position basis is labeled by $\ket{\bar r,\beta}$, where $\beta=(\alpha,\ell)$ combines orbital and layer indices. Wannierization is performed one $d$ at a time, and the resulting coefficients are instead naturally in terms of Wannier orbitals  $\ket{w(\bar R,\beta;d)}$,  carrying the shift $d$ as a label together with $\bar R$ the unit cell and $\beta$ the orbital type.  The two descriptions are related by
\begin{equation}
\ket{w(\bar R,\beta;d)} = \ket{ \bar R+\bar\tau_\alpha +\delta_{\ell,t}d, \beta}.
\end{equation}
Consequently, within the two-center approximation, the Hamiltonian matrix elements in the Wannier orbital basis are directly identified with $\bar{t}_{\beta\beta'}(\delta,d)$ 
\begin{equation}
\bra{w(0,\beta;d)}\Ham_{\rm com}(d) \ket{w(\bar R,\beta';d)} = \bar{t}_{\beta\beta'}(\delta,d),
\end{equation}
where $\delta = \bar R +\bar\tau_{\alpha'} -\bar\tau_\alpha -s_{\ell\ell'}d$ accounts for the true relative position of the Wannier orbitals.

\subsection{Fourier-transform convention}

Throughout this work, we use the Fourier convention
\begin{equation}
W_{\beta\beta'}(\bar k,d) = \sum_{\bar R\in \Lambda} e^{i\bar k\cdot\delta} \bra{w(0,\beta;d)} \Ham_{\rm com}(d) \ket{w(\bar R,\beta';d)},
\label{eq:ourFTConvention}
\end{equation}
in which the phase factor depends on the full displacement $\delta$.
This differs from the convention employed by Wannier90~\cite{Pizzi_2020}, where the phase depends only on the lattice vector $\bar R$. A gauge transformation relates the two conventions, and they both contain identical physical information. The present choice is particularly convenient as it makes the dependence on the relative translation $d$ explicitly periodic over the monolayer lattice,
\begin{equation} \label{eq:periodicity}
W_{\beta\beta'}(\bar k,d+\bar R) = W_{\beta\beta'}(\bar k,d), \quad \bar R\in\Lambda.
\end{equation}
In contrast, the Wannier90 convention adopted in Ref.~\cite{5zt2-scbg} requires an additional phase factor $e^{-i\bar{k}\cdot d\, s_{\ell \ell'}}$ inorder to obtain periodicity in $d$ over $\Lambda$. The resulting pseudo-periodicity of $W_{\beta\beta'}(\bar k,d)$ is therefore not a physical feature, but rather a consequence of the alternative Fourier transform gauge. 

\subsection{Derivation of Eq.~\eqref{eq:DFTtoContinuum}}
Within our conventions, the Bloch Hamiltonian matrix elements $W_{\beta\beta'}(\bar k,d)$ are periodic in $d$ with the periodicity of the monolayer lattice [Eq.~(\ref{eq:periodicity})], and so admit the Fourier expansion
\begin{equation}
W_{\beta\beta'}(\bar k,d)
=
\sum_{\bar G \in \Lambda^*}
e^{-i\bar G\cdot d}
W_{\beta\beta'}(\bar k;\bar G),
\end{equation}
with Fourier coefficients 
\begin{equation} 
W_{\beta\beta'}(\bar{k};\bar G) = \int_{\rm cell} \frac{{\rm d}^2d}{\Acell} \, e^{i\bar G\cdot d} \, W_{\beta\beta'}(\bar{k},d).
\end{equation}
Substituting  the form of $W_{\beta\beta'}(\bar k,d)$ in terms of $\bar{t}_{\beta\beta'}(\delta,d)$ [Eq.~(\ref{eq:WannierBlochCompact})] into the above gives
\begin{equation} 
W_{\beta\beta'}(\bar{k};\bar G) = \int_{\rm cell} \frac{{\rm d}^2d}{\Acell} \, e^{i\bar G\cdot d} \, \sum_{\bar R\in \Lambda} e^{i\bar{k}\cdot\delta}\, \bar t_{\beta\beta'}(\delta,d).
\end{equation}
Combining the above with Eqs.~\eqref{eq:moiretomonolayerperiodicity},~\eqref{eq:moire_FT},~\eqref{eq:registryhyp}, and~\eqref{eq:registryhyp_FT}, we obtain
\begin{equation} \begin{aligned} W_{\beta \beta'}(\bar{k};\bar{G}) &= \int_{\rm cell} \frac{{\rm d}^2d}{\Acell}  \int\frac{d^2q}{\Abz} \sum_{\bar G'\in\Lambda^*} \sum_{\bar R\in \Lambda} e^{-i(q-\bar k)\cdot\delta} \\ &\qquad\times \bar{t}_{\beta\beta'}(q;\bar G') \,e^{-i(\bar{G}'-\bar G)\cdot d}. \end{aligned} \end{equation}
Using $\delta = \bar R +\bar\tau_{\alpha'} -\bar\tau_\alpha -s_{\ell\ell'}d,$ the summation over lattice vectors $\bar{R}$ can be evaluated using the Poisson summation formula 
\begin{equation}
\sum_{\bar R\in \Lambda} e^{-i(q-\bar k)\cdot\bar{R}}=A_{\mathrm{BZ}}\sum_{\bar{G}''}\delta({q-\bar{k}-\bar{G}''}), \end{equation}
enforcing $q=\bar{k}+\bar{G}''$, which collapses the integral over $q$\begin{equation} \begin{aligned} W_{\beta \beta'}(\bar{k};\bar{G}) &= \int_{\rm cell} \frac{{\rm d}^2d}{\Acell}    \sum_{\bar G',\bar{G}''\in\Lambda^*} e^{-i\bar{G}''\cdot(\bar{\tau}_{\alpha'}-\bar{\tau}_{\alpha})} \\ &\qquad\times \bar{t}_{\beta\beta'}(\bar{k}+\bar{G}'';\bar G') \,e^{-i(\bar{G}'-\bar G-\bar{G}'' s_{\ell \ell'})\cdot d}. \end{aligned} \end{equation}
The integral over $d$ imposes the constraint 
\begin{equation}
    \bar{G}'=\bar G+\bar{G}'' s_{\ell \ell'}
\end{equation}
yielding
\begin{equation} \begin{aligned} W_{\beta \beta'}(\bar{k};\bar{G}) &= \sum_{\bar{G}''\in\Lambda^*}e^{i\bar{G}''\cdot(\bar{\tau}_{\alpha}-\bar{\tau}_{\alpha'})}\bar{t}_{\beta\beta'}(\bar{k}+\bar{G}'';\bar G+\bar{G}'' s_{\ell \ell'}). \end{aligned} \end{equation}
Relabeling $\bar{G}'' \rightarrow \bar{H}$, then reproduces  Eq.~(\ref{eq:DFTtoContinuum}).

\end{document}